\documentclass[11pt]{article}
\usepackage[margin=1in]{geometry}

\usepackage{amsmath,amssymb,amsthm}
\usepackage{graphicx}
\usepackage[dvipsnames]{xcolor}
\usepackage{booktabs}
\usepackage{authblk}

\usepackage{mathtools}
\usepackage{bm}

\usepackage{array}
\usepackage{tabularx}
\usepackage{multirow}
\usepackage{listings}

\usepackage[ruled,linesnumbered]{algorithm2e}

\usepackage{tikz}
\usetikzlibrary{positioning,arrows.meta,shapes.geometric,fit,calc,backgrounds,matrix}

\tikzset{
  kgemm/.style    = {rectangle, draw=blue!60!black, fill=blue!12,
                     line width=0.35mm, rounded corners=1pt,
                     inner sep=3pt, text width=38mm, align=center,
                     font=\footnotesize},
  kred/.style     = {rectangle, draw=orange!70!black, fill=orange!25,
                     line width=0.35mm, rounded corners=2pt,
                     inner sep=3pt, text width=38mm, align=center,
                     font=\footnotesize},
  kelem/.style    = {rectangle, draw=black!55, fill=black!6,
                     line width=0.3mm, rounded corners=2pt,
                     inner sep=3pt, text width=38mm, align=center,
                     font=\footnotesize},
  ksample/.style  = {rectangle, draw=violet!60!black, fill=violet!15,
                     line width=0.3mm, rounded corners=2pt,
                     inner sep=3pt, text width=38mm, align=center,
                     font=\footnotesize},
  kcast/.style    = {rectangle, draw=black!55, fill=black!3,
                     line width=0.3mm, dashed,
                     inner sep=3pt, text width=38mm, align=center,
                     font=\footnotesize},
  kloop/.style    = {rectangle, draw=teal!60!black, fill=teal!8,
                     line width=0.35mm, rounded corners=3pt,
                     inner sep=6pt, align=center,
                     font=\footnotesize},
  kgraph/.style   = {rectangle, draw=black!60, dashed, line width=0.35mm,
                     rounded corners=4pt, inner sep=6pt},
  kedge/.style    = {-{Stealth[length=2.2mm,width=1.8mm]},
                     draw=black!75, line width=0.3mm},
  kedgeL/.style   = {-{Stealth[length=2.2mm,width=1.8mm]},
                     draw=black!55, line width=0.28mm},
  legswatch/.style = {rectangle, minimum width=6mm, minimum height=3.5mm,
                      line width=0.3mm},
}

\usepackage{pgfplots}
\usepackage{pgfplotstable}
\pgfplotsset{compat=1.18}
\usepgfplotslibrary{fillbetween,groupplots}
\usepackage{subcaption}

\providecommand{\plotsdir}{experiments/plots}
\providecommand{\resultsdir}{experiments/results}

\IfFileExists{experiments/plots/pgfplots-style.tex}{
\pgfplotsset{compat=1.18}

\definecolor{ColA}{HTML}{1F5582}
\definecolor{ColC}{HTML}{9E2A2B}

\definecolor{BinSGD}{HTML}{9E2A2B}
\definecolor{BinSGDBF}{HTML}{E89090}
\definecolor{BinCA}{HTML}{1F5582}
\definecolor{BinCATF}{HTML}{3E7CB1}
\definecolor{BinCABF}{HTML}{7DB7D9}
\definecolor{BinCABFD}{HTML}{ABC7DA}
\definecolor{BinCAFP16}{HTML}{5B9E9E}

\pgfplotsset{
  validation/.style={
    width=8.5cm, height=6cm,
    grid=major, tick align=outside,
    every axis legend/.append style={font=\small, draw=none, fill=none},
    label style={font=\small},
    tick label style={font=\footnotesize},
    title style={font=\small},
  }
}

\pgfplotsset{
  scaling/.style={
    width=8.5cm, height=6cm,
    grid=major, tick align=outside,
    every axis legend/.append style={font=\small, draw=none, fill=none},
    label style={font=\small},
    tick label style={font=\footnotesize},
    title style={font=\small},
    cycle list={
      {BinSGD,    mark=*,         thick, mark size=2pt},
      {BinSGDBF,  mark=triangle*, thick, mark size=2.5pt},
      {BinCA,     mark=square*,   thick, mark size=2pt},
      {BinCATF,   mark=square,    thick, mark size=2pt,   dashed},
      {BinCABF,   mark=diamond*,  thick, mark size=2.5pt},
      {BinCABFD,  mark=diamond,   thick, mark size=2.5pt, dashed},
      {BinCAFP16, mark=asterisk,  thick, mark size=2.5pt, dotted},
    },
  }
}
}{}

\usepackage[numbers,sort&compress]{natbib}
\usepackage{hyperref}

\usepackage[capitalize,noabbrev]{cleveref}

\providecommand{\Description}[1]{}

\usepackage{aliascnt}
\newcommand{\newtheoremshared}[3]{%
  \newaliascnt{#1}{theorem}%
  \newtheorem{#1}[#1]{#2}%
  \aliascntresetthe{#1}%
  \crefname{#1}{#2}{#3}%
  \Crefname{#1}{#2}{#3}%
}
\theoremstyle{plain}
\newtheorem{theorem}{Theorem}[section]
\newtheoremshared{lemma}{Lemma}{Lemmas}
\newtheoremshared{proposition}{Proposition}{Propositions}
\newtheoremshared{corollary}{Corollary}{Corollaries}

\theoremstyle{definition}
\newtheoremshared{definition}{Definition}{Definitions}
\newtheoremshared{assumption}{Assumption}{Assumptions}
\newtheoremshared{example}{Example}{Examples}

\theoremstyle{remark}
\newtheoremshared{remark}{Remark}{Remarks}

\crefname{theorem}{Theorem}{Theorems}
\Crefname{theorem}{Theorem}{Theorems}

\newcommand{\R}{\mathbb{R}}
\newcommand{\E}{\mathbb{E}}

\newcommand{\Frob}{\mathrm{F}}
\newcommand{\norm}[1]{\left\lVert #1 \right\rVert}
\newcommand{\fnorm}[1]{\left\lVert #1 \right\rVert_{\Frob}}
\newcommand{\iterhat}[1]{\widehat{x}^{(#1)}}
\newcommand{\bigO}{\mathcal{O}}
\newcommand{\abs}[1]{\left\lvert #1 \right\rvert}

\lstdefinestyle{cpp}{ basicstyle=\ttfamily\footnotesize, keywordstyle=\color{blue!60!black}\bfseries, commentstyle=\color{green!40!black}\itshape, stringstyle=\color{red!60!black}, language=C++, breaklines=true, frame=single, framerule=0pt, backgroundcolor=\color{black!3}, showstringspaces=false, tabsize=2, } 

\title{Mixed-Precision Communication-Avoiding SGD for Generalized Linear Models on GPUs}

\author[1]{Aditya Devarakonda}
\author[2]{Irene Sim\'{o} Mu\~{n}oz}
\author[2]{Giulia Guidi}
\affil[1]{Department of Computer Science, Wake Forest University, Winston-Salem, NC, USA\\ \texttt{devaraa@wfu.edu}}
\affil[2]{Department of Computer Science, Cornell University, Ithaca, NY, USA\\ \texttt{\{is449,\,gguidi\}@cornell.edu}}
\date{}

\begin{document}

\maketitle

\begin{abstract}
Distributed stochastic gradient descent (SGD) is limited by communication rather than computation, since each iteration requires an AllReduce across processes.
Communication-avoiding SGD (CA-SGD) amortizes communication over $s$ iterations by replacing $s$ consecutive AllReduces with a single AllReduce of an $sb\times sb$ Gram matrix, trading more computation and bandwidth for fewer synchronization points.
Modern GPUs with matrix hardware and reduced-precision formats offset this by accelerating the Gram GEMM and shrinking BF16 traffic.
We study mixed-precision CA-SGD for generalized linear models on NVIDIA GPUs.
Our finite-precision analysis decomposes the local rounding error of one CA-SGD outer iteration into nine independent precision choices, depending on the hardware only through its low-precision unit roundoffs, so the resulting recipes transfer in principle across GPU generations.
The recipe stores the input matrix and margin vector in low precision, computes the Gram matrix from low-precision inputs with high-precision accumulation, communicates it in high precision, and performs the inner recurrence and weight updates in high precision.
On NERSC Perlmutter A100 GPUs, mixed-precision CA-SGD matches FP32 SGD loss within $0.5\%$ on logistic, linear, and Poisson problems and reaches $5.1$--$6.8\times$ speedup over FP32 SGD on \texttt{epsilon}, \texttt{SUSY}, \texttt{HIGGS}, \texttt{synth}, and \texttt{Poisson-synth}.
Our software is available at \url{https://doi.org/10.5281/zenodo.20448273}
\end{abstract}

\medskip
\noindent\textbf{Keywords:} Communication-avoiding algorithms, mixed precision, stochastic
gradient descent, generalized linear models, GPU, NCCL, BF16

\section{Introduction}
\label{sec:intro}

Distributed mini-batch stochastic gradient descent (SGD) is one of the most widely used optimization methods for large-scale empirical risk minimization.
Each iteration of SGD requires a synchronizing AllReduce across processes to aggregate partial gradients into a single solution update vector~\cite{bottou2018optimization}.
In the 1D block-column setting used here, the AllReduce synchronization costs $2\alpha\log P + \beta\,W \cdot 2(P-1)/P$, where the latency term $\alpha\log P$ dominates for the small message sizes typical of mini-batch SGD~\cite{chan2007collective,thakur2005collective}.
Because local computation scales with $P$ but communication latency does not, scalable mini-batch SGD requires algorithmic changes that reduce synchronization frequency.

\paragraph{Communication-avoiding SGD}

Communication-avoiding SGD (CA-SGD) addresses this gap by amortizing communication over $s$ consecutive mini-batch iterations~\cite{devarakonda2020ca, devarakonda2025hybridsgd}.
Rather than performing $s$ separate vector AllReduces of length $b$, CA-SGD samples a single \emph{outer block} of $sb$ rows ($Y \in \R^{sb \times n_{\mathrm{loc}}}$), where $n_{\mathrm{loc}}$ is the length of the locally stored features.
CA-SGD then computes the local margin contribution $r = Yx \in \mathbb{R}^{sb}$ and the local Gram matrix $G = YY^{\!\top} \in \mathbb{R}^{sb \times sb}$, after which a single grouped AllReduce is performed to obtain the global margin and Gram matrix.
CA-SGD then performs $s$ inner residual-correction updates independently and redundantly on each process without further communication.
The reformulation trades fewer synchronization points for a larger Gram GEMM and a larger grouped AllReduce.
On CPU implementations, this trade-off limits the useful range of $s$ because the Gram GEMM and grouped AllReduce payload grow as $(sb)^2$.
Tensor-core GPUs change this balance by accelerating the Gram GEMM and reducing traffic with low-precision storage.
The performance opportunity does not imply a uniform BF16 recipe, however, because the outer iteration combines operations with different rounding-error sensitivities.

A single outer CA-SGD iteration involves nine distinct operations (\Cref{subsec:notation-nine-u}) whose rounding-error tolerances vary by orders of magnitude: the Gram GEMM and margin GEMV are length-$n_{\mathrm{loc}}$ inner products whose accumulator precision sets the leading error, the AllReduced quantities carry a collective factor $C(P)$ set by the reduction topology and summation order, and the inner correction sum of length up to $(s{-}1)b$ requires $qu<1$ for the Higham factor and therefore FP32 storage.
Uniform low precision provides insufficient precision for error-sensitive operations.

Our work adapts this finite-precision program to stochastic $s$-step methods by bounding the local error of one CA-SGD outer iteration and lifting the resulting perturbation into a bounded-regime convergence statement.
We focus on generalized linear models, including linear, logistic, and Poisson regression in the empirical study.
Because Poisson regression has an unbounded inverse-link derivative, the theorem-covered loss family is the Lipschitz-residual subset instantiated by logistic and linear regression.

\paragraph{Contributions.}

This paper makes four contributions.
(C1) It presents the first finite-precision analysis of CA-SGD (\Cref{thm:mixed-precision}), where the local forward error of one outer iteration decomposes into nine additive terms.
Each term depends on a single precision choice and can be computed from the algorithmic parameters and the bounds in \Cref{ass:norm,ass:lipschitz,ass:gram-bound,ass:allreduce,ass:weight}.
The decomposition exposes a constraint: the inner $s$-step correction sum must be performed in FP32, which a uniform low-precision choice would overlook.
(C2) A concrete A100 mixed-precision recipe (\Cref{tab:recipe}) requires BF16 storage and BF16-input/FP32-accumulate tensor-core math for the Gram GEMM, margin, and outer-gradient GEMVs. It retains FP32 for the inner correction, residual, master weights, and Gram AllReduce, and uses BF16 for the margin AllReduce.
The recipe is based directly on the nine per-choice coefficients $\alpha_{\ast}$ rather than selected through experimentation.
(C3) A perturbed-SGD convergence guarantee (\Cref{thm:global-perturbed}) gives, in a bounded regime with local strong convexity and smoothness, a deterministic radius $\rho$ around the exact SGD trajectory.
(C4) In empirical validation on NERSC Perlmutter A100 GPUs with up to $P=256$ GPUs, Recipe~C matches the FP32 SGD baseline within $0.5\%$ and reaches up to ${\sim}16\times$ speedup over FP32 SGD on \texttt{synth} at $(b,s)=(1,256)$ and $5.2$--$6.2\times$ speedup over FP32 SGD on \texttt{epsilon}, \texttt{SUSY}, \texttt{HIGGS}, \texttt{synth}, and \texttt{Poisson-synth} at $(b,s)=(8,64)$.
The Recipe~D variant, which casts only the Gram AllReduce to BF16, reaches up to $6.8\times$ on these datasets but trades a larger relative loss gap for the speedup and falls outside the deterministic budget that the theory verifies for Recipe~C.
The Recipe~C Gram GEMM reaches $92\%$ of the A100 BF16 tensor-core peak.

\section{Background and Related Work}
\label{sec:background}

\subsection{Generalized linear models and distributed mini-batch SGD}
\label{subsec:bg-glm-sgd}

A generalized linear model (GLM) for features $A\in\R^{m\times n}$ and labels $y$ minimizes the empirical risk $F(x)=m^{-1}\sum_{i=1}^{m}L(a_i^{\!\top}x,y_i)$, with $L$ depending on the weights only through the margin $r_i=a_i^{\!\top}x$~\cite{mccullagh1989glm}.
We focus on three canonical losses: logistic ($L(r,y)=\log(1+e^{-yr})$, $L_{\sigma}=1/4$), linear ($L(r,y)=(r-y)^{2}/2$, $L_{\sigma}=1$), and Poisson ($L(r,y)=e^{r}-yr$, $L_{\sigma}$ unbounded).
The gradient is $\nabla F(x) = -m^{-1}A^{\!\top}\delta$, where the residual is $\delta_i = -\partial L/\partial r\,(r_i, y_i)$. Both logistic and linear models have Lipschitz residuals. The logistic residual is globally bounded, and the linear residual is bounded in the bounded-margin regimes used in \Cref{ass:lipschitz}.

Distributed mini-batch SGD samples $I \subseteq \{1, \dots, m\}$ of size $b$ and computes $g_I(x) = -b^{-1} Y_I^{\!\top} \delta( Y_Ix, y_I)$. 
Under the bounded-trajectory, local-curvature model of \Cref{ass:weight,ass:global-opt} and for $\eta \le 1/L_F$, $g_I$ is unbiased with variance $\sigma_g^2 / b$, and the iterate satisfies the canonical contraction~\cite{bottou2018optimization,ajalloeian2020biased}.
In this work, we use 1D block-column partitioning ($n_{\mathrm{loc}} = n / P$ columns per rank, \Cref{sec:comm-cost}), which requires a single vector AllReduce of length $b$ for each mini-batch step.
The 2D and hybrid variants of~\cite{devarakonda2025hybridsgd,wang2025eccd} further reduce communication volume but still require one synchronization step per mini-batch and are complementary to the $s$-step amortization used here.

\subsection{Communication-avoiding and $s$-step methods}
\label{subsec:bg-cassgd}

The $s$-step approach~\cite{demmel1993parallel,hoemmen2010thesis,carson2015thesis,demmel2012communication} replaces $s$ communication-blocking linear algebra primitives with a single grouped collective on a larger composite object, followed by $s$ local steps of the original recurrence. 
CA-SGD~\cite{devarakonda2020ca,devarakonda2025hybridsgd,wang2025eccd} applies this to mini-batch SGD: sample $sb$ rows, compute the rank-$(sb)$ Gram matrix $G = YY^{\!\top}$ and the base margin $r = Yx$ in parallel, fuse them into one grouped AllReduce, then run $s$ local residual-correction updates $z^{(j)} = r_j + (\eta/b)\,G_{j,<j}\,\delta_{<j}$ on the inner iterate before forming the outer gradient $g = Y^{\!\top}\delta_{\mathrm{hist}}$ and updating the weights.
The per-block cost (\Cref{tab:abg-comparison}) replaces $s$ vector AllReduces with one grouped AllReduce, at the cost of a Gram GEMM, an $(sb)^2$-word payload, and an $\bigO(b^2 s^2)$ inner correction sum.
On CPUs, the $\gamma(sb)^2 n_{\mathrm{loc}}$ term breaks the trade-off at modest $s$. 
On tensor-core GPUs, this term is suppressed by up to $16\times$ in compute peak and $2\times$ in traffic, so the crossover shifts to smaller $s$ (\Cref{tab:expt-d,fig:roofline-combined}).

\paragraph{$s$-step error analysis}

The numerical stability of $s$-step Krylov methods has been studied extensively. The residual-replacement analysis of $s$-step CG and BiCG~\cite{carson2014residualreplacement} identifies the basis condition number $\kappa(V_s)$ as the main amplifier of the gap between the recursively updated and true residual.
This finite-precision analysis of classical Lanczos was extended to $s$-step Lanczos~\cite{carson2015sstep} and led to adaptive techniques~\cite{carson2018adaptive} that stabilize the algorithm at runtime.
Recent work by Carson, Gergelits, and Yamazaki~\cite{carson2022sstepmixed} extends prior work to non-uniform floating-point storage choices within the same error analysis framework.
Here, we extend this line of finite-precision analysis to $s$-step SGD by bounding the per-outer-iteration error and proving that the resulting perturbation remains within the bounded-regime SGD convergence neighborhood of \Cref{thm:global-perturbed}.

\subsection{Mixed-precision arithmetic and finite-precision analysis on modern GPUs}
\label{subsec:bg-mixed}

The NVIDIA A100 exposes FP32, TF32, FP16, and BF16 with unit roundoffs $\approx 6\!\times\!10^{-8}$, $5\!\times\!10^{-4}$, $5\!\times\!10^{-4}$, and $4\!\times\!10^{-3}$, respectively~\cite{nvidia2021a100}.
TF32 and BF16 share the FP32 exponent range but have truncated mantissas, while FP16 uses the narrower IEEE half-precision range.
Tensor cores accept BF16 or FP16 inputs with FP32 accumulation at $312$~TFLOP/s for both formats, and A100 also supports FP16-input/FP16-accumulate HMMA for Recipe~F.
The dominant approach uses low-precision inputs with higher-precision accumulation, with Higham factor $2u_{16}+\gamma_n^{(u_{32})}$ (\Cref{eq:bf16-fp32-accum}) in the BF16-input/FP32-accumulate rows of \Cref{thm:mixed-precision}.

The deterministic Higham factor $\gamma_q^{(u)} = qu/(1 - qu)$ used throughout \Cref{sec:analysis} is standard~\cite{higham2002accuracy,golub2013matrix}. The Higham--Mary probabilistic refinement~\cite{higham2019probabilistic,higham2020probabilistic} would substitute $\lambda(\rho)\,u\sqrt{q}$, but this is left for future work.
The closest stylistic precedents for our recipe table (\Cref{tab:recipe}) are the three-precision iterative-refinement analyses by Carson and Higham~\cite{carson2018new,carson2018accelerating,carson2020threeprecision}. We have nine slots instead of three because the CA-SGD outer iteration includes more kernel categories.
A parallel study on GPU tensor-core mixed precision for dense linear system solvers~\cite{haidar2018tensorcore, carson2018accelerating} shows that aggressive low-precision factorization preserves FP64 backward stability when the refinement loop is well-conditioned.
Higham et al.~\cite{higham2019simulating} provide a software simulation methodology to inform effective roundoff configurations.

\subsection{Mixed-precision machine-learning training}
\label{subsec:bg-mp-ml}

Mixed precision in machine learning has primarily been motivated by deep learning.
Micikevicius et al.~\cite{micikevicius2018mixed} introduced FP32 master weights with FP16 forward and backward computations and loss scaling. Kalamkar et al.~\cite{kalamkar2019bf16} showed that BF16's wider range can eliminate loss scaling in most training runs.
Both approaches have been integrated into production software and frameworks such as PyTorch \texttt{AMP}, NVIDIA Apex, and cuBLAS \texttt{Tensor\_Op}.
This work develops a finite-precision analysis for CA-SGD and uses it to select a mixed-precision recipe, with its accuracy and speedup measured in \Cref{sec:experiments}.

\section{Notation and Finite-Precision Model}
\label{sec:notation}

We fix the notation, finite-precision model, and storage-precision mapping that the algorithms of \Cref{sec:algorithm-design} and the error bounds of \Cref{sec:analysis} both depend on.

\subsection{Algorithmic quantities}
\label{subsec:notation-algorithmic}

Let $A\in\R^{m\times n}$ be the row-standardized data matrix and $y\in\R^m$ the labels ($\{-1,+1\}^m$ for logistic, $\R^m$ for linear, $\mathbb{Z}_{\ge 0}^m$ for Poisson).
Let $Y=SA\in\R^{sb\times n}$ be the $sb$-row subsampled matrix drawn during one outer iteration, where $S\in\{0,1\}^{sb\times m}$ is a sampling matrix with exactly one non-zero per row that selects a row of $A$.
Rows of $A$ may be selected multiple times among the $sb$ outer samples (i.i.d.\ uniform sampling with replacement, see \Cref{ass:global-opt} and \Cref{alg:mp-ca-sgd}), so a column of $S$ may contain zero, one, or several nonzeros.
Let $y_I = Sy \in \R^{sb}$ be the corresponding label vector, and $x \in \R^n$ the weight iterate, with $x^{(h)}$ denoting its value after $h$ outer iterations.
The CA-SGD recurrence also maintains the Gram block $G = YY^\top \in \R^{sb \times sb}$, the base margin $r = Yx \in \R^{sb}$, the per-sample residual $\delta(r, y) = -\partial L / \partial r \in \R^{sb}$ (the negative loss gradient with respect to the margin), and an inner iterate that propagates through $s$ residual corrections per outer iteration.
The quantity $g:=Y^\top\delta\in\R^n$ assembled by the outer gradient GEMV is the (sample-summed) \emph{descent direction}. The sample-mean true gradient estimator used in \Cref{ass:global-opt,thm:global-perturbed} is $\nabla_x\!\widehat L(x)=-b^{-1}Y_I^\top\delta=-b^{-1}g_I$, and the weight update $\widetilde x \leftarrow \widetilde x+(\eta/b)\,g$ in \Cref{alg:mp-ca-sgd} is the standard SGD step $x\leftarrow x-\eta\,\nabla_x\!\widehat L(x)$ written in $\delta$-form.
We use the unqualified term \emph{gradient} to refer to $g$ when discussing the kernel and its precision slot $u_g$ (defined in \Cref{subsec:notation-nine-u}), with the understanding that the sign and $b^{-1}$ normalization are absorbed into the step size. 
We refer to $\delta$ as the \emph{residual}, since $\delta$ is an intermediate per-sample quantity in $\R^{sb}$ whose storage precision is tracked separately ($u_r$, $u_\sigma$, and $u_c$).
The residual is loss-specific with the following definitions: $\delta_i=y_i(1-\sigma(y_i r_i))$ for logistic regression ($L_\sigma=1/4$), $\delta_i=y_i-r_i$ for linear regression ($L_\sigma=1$), and $\delta_i=y_i-\exp(r_i)$ for Poisson regression.
Poisson regression has an unbounded Lipschitz value ($L_\sigma=\infty$), which places this loss outside the analysis. We report its empirical behavior as boundary-stress evidence.
We use $b$ for the mini-batch size, $s$ for the CA inner-loop length, $P$ for the number of ranks (e.g., GPUs or CPUs), and $n_{\rm loc}=n/P$ for the number of local features stored per rank.

\subsection{Finite-precision model}
\label{subsec:notation-finite-precision}
\label{subsec:notation-prob-rounding}

Here, we work under the standard IEEE-754 rounding model: every elementary floating-point operation $\mathrm{op}\in\{+,-,\times,\div\}$ evaluated in a format with unit roundoff $u$ satisfies $\mathrm{fl}_u(x\ \mathrm{op}\ y)=(x\ \mathrm{op}\ y)(1+\theta)$ with $\abs{\theta}\le u$~\cite{higham2002accuracy}.
For a stored vector or matrix $z$ in unit-roundoff $u_\ast$, we write $\widetilde z\in\mathrm{fl}_{u_\ast}(\R^{\cdot})$ for $\widetilde z=z+e$ with $\abs{e}\le u_\ast\abs{z}$ componentwise, which yields $\norm{e}\le u_\ast\norm{z}$ for any vector $p$-norm or matrix Frobenius norm.
Composite kernels (GEMV, GEMM, AllReduce, sigmoid) inherit Higham-style entrywise bounds that add to $\bigO(u_\ast)$, and second-order terms $\bigO(u_\ast^2)$ are dropped.

The analysis in \Cref{sec:analysis} accumulates rounding error along inner-product reductions of various lengths $q$: the margin and Gram kernels with $q=n_{\mathrm{loc}}$, the outer gradient with $q=sb$, and the inner correction block at iteration $j$ with $q=jb$.
The standard deterministic Higham bound for the floating-point inner product of two length-$q$ vectors $x,y$ accumulated in precision $u$ is
\begin{equation}
\label{eq:notation-gamma-q}
  \abs{\mathrm{fl}_u(x^{\!\top}\!y) - x^{\!\top}\!y}
  \;\le\;
  \gamma_q^{(u)}\,\abs{x}^{\!\top}\!\abs{y},
  \qquad
  \gamma_q^{(u)} := \frac{q\,u}{1-q\,u}
  \;=\; q\,u + \bigO((qu)^{2}),
\end{equation}
valid whenever $qu<1$~\cite{higham2002accuracy}.
We use $\gamma_q^{(u)}$ throughout \Cref{sec:analysis} for all GEMV/GEMM reductions.
The AllReduce constant $C(P)$ of \Cref{ass:allreduce} is an empirical hardware parameter depending on the reduction order and routing algorithm (e.g.\ tree vs.\ ring). It appears as $C(P)\,u_\ast$ when bounding error associated with communicated quantities.
For low-precision-input/FP32-accumulate kernels we replace $\gamma_q^{(u)}$ with the input-cast plus FP32-accumulation factor. The BF16 specialization is \Cref{eq:bf16-fp32-accum}.
The Higham--Mary probabilistic-rounding model~\cite{higham2019probabilistic,higham2020probabilistic} would substitute $\lambda(\rho)\,u_\ast\sqrt q$ for the deterministic $qu$ factor, which we leave for future work.

\subsection{The nine unit roundoffs}
\label{subsec:notation-nine-u}

A single outer iteration of CA-SGD executes nine distinct operations, each with its precision tracked separately.
The first-order analysis of \Cref{thm:mixed-precision} separates the leading finite-precision error among these nine precision choices, allowing each precision to be independently selected based on the desired error budget.
The nine symbols are $u_{A}$ for storage of $A$ (and of the gathered subsampled matrix $Y$), $u_{G}$ for the Gram GEMM $G = Y Y^{\top}$, $u_{r}$ for the base margin GEMV $r = Y x$, $u_{c}$ for the inner-loop correction sum (the CA-SGD correction recurrence), $u_{\sigma}$ for the elementwise nonlinearity kernel (e.g., sigmoid for logistic regression), $u_{g}$ for the outer gradient GEMV $g = Y^{\top} \delta$, $u_{\mathrm{AR},r}$ for the margin AllReduce datatype, $u_{\mathrm{AR},G}$ for the Gram AllReduce datatype, and $u_{x}$ for weight storage and update.
These nine subscripts appear in the algorithms of \Cref{sec:algorithm-design} and in every error term $E_\ast$ of \Cref{sec:analysis}.

\subsection{Storage-precision mapping}
\label{subsec:notation-storage-map}

\Cref{tab:storage-map} lists the quantities referenced by the algorithm floats of \Cref{sec:algorithm-design}, along with their storage-precision variable. 
We use $\widetilde z$ for the finite-precision quantity associated with an exact-arithmetic quantity $z$. For example, $\widetilde Y \in \mathrm{fl}_{u_A}(\R^{sb \times n})$ is the finite-precision version (defined by $u_A$) of the sampled submatrix $Y$.
Wherever an algorithm or theorem statement uses $\widetilde\cdot$, the storage precision is the one listed in \Cref{tab:storage-map}.

\begin{table}[ht]
\centering
\caption{Storage-precision mapping for the quantities in one outer CA-SGD iteration.
The middle column gives the storage-precision variable. The right column gives the concrete A100 realization in our experiments.
Labels $y$ are stored independently of $A$ at FP32 throughout.}
\label{tab:storage-map}
\small
\begin{tabular}{@{}llll@{}}
\toprule
Symbol & Quantity & Storage precision & Typical format \\
\midrule
$\widetilde{A}$, $\widetilde{Y}$ & feature matrix, subsampled rows & $u_{A}$ & BF16 \\
$\widetilde{G}$                 & Gram block $G = Y Y^{\top}$   & $u_{G}$ & BF16 in; FP32 accum/out \\
$\widetilde{r}$               & base margin $Y x$             & $u_{r}$ & BF16 in; FP32 accum; BF16 out \\
$\widetilde{\delta}$            & residual      & $u_{c}$, $u_{\sigma}$ & FP32 \\
$y$, $y_{I}$                    & labels, sampled labels        & FP32 & FP32 \\
$\widetilde{g}$                 & outer gradient $Y^{\top}\delta$ & $u_{g}$ & BF16 in; FP32 accum/out \\
$\widetilde{x}$                 & weights                & $u_{x}$   & FP32 \\
-                          & margin / Gram AllReduce       & $u_{\mathrm{AR},r}$, $u_{\mathrm{AR},G}$ & BF16, FP32 \\
\bottomrule
\end{tabular}
\end{table}

\section{Algorithm Design}
\label{sec:algorithm-design}
We present the mixed-precision CA-SGD outer iteration that the analysis of \Cref{sec:analysis} tracks one precision slot at a time, then state its communication cost and place it on the roofline.
The FP32 baseline is the special case of the algorithm in which every precision slot $u_{\ast}$ of \Cref{tab:storage-map} is set to $u_{\rm FP32}$.

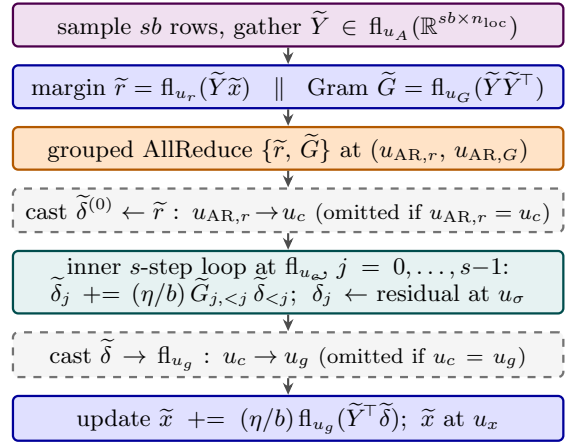
\begin{figure}[t]
  \centering
  \begin{tikzpicture}[
  font=\footnotesize,
  node distance=2.2mm,
  every node/.style={align=center, rounded corners=2pt, draw, line width=0.3mm,
                     inner xsep=4pt, inner ysep=2.5pt, text width=70mm},
  sample/.style ={fill=violet!12, draw=violet!60!black},
  comp/.style   ={fill=blue!12,   draw=blue!60!black},
  red/.style    ={fill=orange!22, draw=orange!70!black},
  loop/.style   ={fill=teal!10,   draw=teal!60!black, text width=70mm},
  cast/.style   ={fill=black!4,   draw=black!55, dashed, text width=70mm},
  arr/.style    ={-{Stealth[length=1.6mm,width=1.4mm]},
                  draw=black!70, line width=0.28mm},
]
  \node[sample]                  (samp) {sample $sb$ rows, gather $\widetilde Y \in \mathrm{fl}_{u_A}(\R^{sb\times n_{\rm loc}})$};
  \node[comp, below=of samp]     (gem)  {margin $\widetilde r = \mathrm{fl}_{u_r}(\widetilde Y \widetilde x)$ \;\;$\parallel$\;\; Gram $\widetilde G = \mathrm{fl}_{u_G}(\widetilde Y \widetilde Y^{\!\top})$};
  \node[red,  below=of gem]      (ar)   {grouped AllReduce $\{\widetilde r,\,\widetilde G\}$ at $(u_{\mathrm{AR},r},\,u_{\mathrm{AR},G})$};
  \node[cast, below=of ar]       (prom) {cast $\widetilde\delta^{(0)} \leftarrow \widetilde r$ : $u_{\mathrm{AR},r}\!\to\!u_c$ \scriptsize(omitted if $u_{\mathrm{AR},r}=u_c$)};
  \node[loop, below=of prom]     (loop) {inner $s$-step loop at $\mathrm{fl}_{u_c}$, $j=0,\dots,s{-}1$:\\[-1pt]
                                          $\widetilde\delta_j \mathrel{+}= (\eta/b)\,\widetilde G_{j,<j}\,\widetilde\delta_{<j}$;\;
                                          $\widetilde\delta_j \leftarrow$ residual at $u_\sigma$};
  \node[cast, below=of loop]     (dem)  {cast $\widetilde\delta \to \mathrm{fl}_{u_g}$ : $u_c\!\to\!u_g$ \scriptsize(omitted if $u_c=u_g$)};
  \node[comp, below=of dem]      (upd)  {update $\widetilde x \mathrel{+}= (\eta/b)\, \mathrm{fl}_{u_g}(\widetilde Y^{\!\top}\widetilde\delta)$;\;\;$\widetilde x$ at $u_x$};

  \draw[arr] (samp) -- (gem);
  \draw[arr] (gem)  -- (ar);
  \draw[arr] (ar)   -- (prom);
  \draw[arr] (prom) -- (loop);
  \draw[arr] (loop) -- (dem);
  \draw[arr] (dem)  -- (upd);
\end{tikzpicture}
  \caption[Mixed-precision CA-SGD kernel DAG]{Mixed-precision
  CA-SGD for one outer iteration on each rank.
  Each kernel and storage box is labeled by its precision slot from
  \Cref{tab:storage-map}. Dashed boxes are precision casts before
  and after the inner $s$-step loop (omitted when input/output
  precisions agree).
  Recipe~C (\Cref{tab:recipe}) instantiates
  $u_A=u_r=u_G=u_{{\rm AR},r}=u_g=u_{\rm BF16}$ and
  $u_c=u_\sigma=u_{{\rm AR},G}=u_x=u_{\rm FP32}$.}
  \Description{Directed acyclic graph of one outer iteration of mixed-precision CA-SGD: a sample-and-gather step at storage precision u_A, parallel margin GEMV at u_r and Gram GEMM at u_G, a grouped AllReduce at split precisions (u_AR_r, u_AR_G), a precision-cast box from u_AR_r to u_c, the inner s-step correction-and-residual loop at u_c with residual at u_sigma, a precision-cast box from u_c to u_g, and a final gradient GEMV / weight update at u_g into master weights at u_x.}
  \label{fig:dag-casgd-mp}
\end{figure}

\subsection{Mixed-precision algorithm}
\label{subsec:alg-floats}

\Cref{alg:mp-ca-sgd} states the mixed-precision CA-SGD outer iteration.
Every assignment that rounds in finite precision is wrapped in $\mathrm{fl}_{u_{\ast}}(\cdot)$, where the subscript $u_{\ast}$ names the slot that carries the rounding error.
Each numbered line corresponds to one of the nine $E_{\ast}$ terms that appear in \Cref{thm:mixed-precision}.

\begin{algorithm}[t]
\DontPrintSemicolon
\SetKwInput{KwData}{Input}
\SetKwInput{KwResult}{Output}
\KwData{%
  $\widetilde{A}_{p} \in \mathrm{fl}_{u_{A}}(\R^{m \times n_{\text{loc}}})$
  (rank-$p$ block-column feature slab, BF16);
  labels $y \in \R^{m}$ stored in FP32;
  weights $\widetilde{x} \in \mathrm{fl}_{u_{x}}(\R^{n_{\text{loc}}})$;
  step size $\eta$, batch size $b$, CA parameter $s$.}
\KwResult{Updated $\widetilde{x} \in \mathrm{fl}_{u_{x}}(\R^{n_{\text{loc}}})$.}
\BlankLine
Sample $sb$ batch indices $I = I_{0} \cup \cdots \cup I_{s-1}$, $\abs{I} = sb$, drawn i.i.d.\ uniformly with replacement from $\{0,\dots,m-1\}$\;
Gather sampled rows $\widetilde{Y}_{I} \in \mathrm{fl}_{u_{A}}(\R^{sb \times n_{\text{loc}}})$\;
$\widetilde{r}_{\text{loc}} \leftarrow \mathrm{fl}_{u_{r}}\!\bigl(\widetilde{Y}_{I}\, \widetilde{x}\bigr)$ \tcp*{base margin GEMV; $u_{A} \to u_{r}$}
$\widetilde{G}_{\text{loc}} \leftarrow \mathrm{fl}_{u_{G}}\!\bigl(\widetilde{Y}_{I}\, \widetilde{Y}_{I}^{\top}\bigr)$ \tcp*{Gram GEMM; $u_{A} \to u_{G}$}
$\bigl[\widetilde{r};\, \widetilde{G}\bigr] \leftarrow \mathrm{AllReduce}\Bigl(\bigl[\widetilde{r}_{\text{loc}};\, \widetilde{G}_{\text{loc}}\bigr];\ \bigl(u_{\mathrm{AR},r},\, u_{\mathrm{AR},G}\bigr)\Bigr)$ \tcp*{grouped NCCL, split datatype}
$\widetilde{\delta}_{\text{hist}} \leftarrow [\,]$\;
\For{$j \leftarrow 0$ \KwTo $s - 1$}{
  \uIf{$j = 0$}{
    $\widetilde{z}^{(0)} \leftarrow \widetilde{r}_{0}$\;
  }
  \Else{
    $\widetilde{z}^{(j)} \leftarrow \mathrm{fl}_{u_{c}}\!\bigl(\widetilde{r}_{j} + \tfrac{\eta}{b}\, \widetilde{G}_{j,<j}\, \widetilde{\delta}_{\text{hist}}\bigr)$ \nllabel{line:correction}\tcp*{corrected margin at $u_c$}
  }
  $\widetilde{\delta}^{(j)} \leftarrow \mathrm{fl}_{u_{\sigma}}\!\bigl(\delta(\widetilde{z}^{(j)}, y_{I_{j}})\bigr)$ \nllabel{line:residual}\tcp*{per-batch residual at $u_\sigma$}
  $\widetilde{\delta}_{\text{hist}} \leftarrow \bigl[\widetilde{\delta}_{\text{hist}};\, \widetilde{\delta}^{(j)}\bigr]$ \nllabel{line:hist}\;
}
$\widetilde{g}_{\text{loc}} \leftarrow \mathrm{fl}_{u_{g}}\!\bigl(\widetilde{Y}_{I}^{\top}\, \widetilde{\delta}_{\text{hist}}\bigr)$ \nllabel{line:gradient}\tcp*{outer gradient GEMV}
$\widetilde{x} \leftarrow \mathrm{fl}_{u_{x}}\!\Bigl(\widetilde{x} + \tfrac{\eta}{b}\, \widetilde{g}_{\text{loc}}\Bigr)$ \nllabel{line:update}\tcp*{weight update}
\caption{Mixed-precision CA-SGD outer iteration.}
\label{alg:mp-ca-sgd}
\end{algorithm}

\subsection{Communication cost and roofline}
\label{sec:comm-cost}
\label{sec:perf-model}

To model communication cost, we use the standard $\alpha$-$\beta$-$\gamma$ model for parallel runtime.
Per rank, the runtime is $T = \gamma F + \beta W + \alpha L$, where $F$ is the number of flops, $W$ is the number of words, and $L$ is the number of messages.
In this model, mixed precision is used in two places.
The word size is reduced from FP32 to BF16, resulting in $\beta_{\rm BF16} = \beta_{\rm FP32} / 2$.
The compute constant also decreases, with $\gamma_{\rm BF16}$ up to $16\times$ smaller than $\gamma_{\rm FP32}$ on tensor cores, and we assume 1D block-column partitioning across $P$ ranks, with $n_{\text{loc}} = n/P$ columns per rank.
The AllReduce cost under this partitioning is
\begin{equation}
\label{eq:allreduce-cost}
  L_{\mathrm{AR}}(W) = 2\alpha \log P + \beta\, W \cdot \tfrac{2(P-1)}{P}.
\end{equation}
\Cref{tab:abg-comparison} compares the cost of $s$ classical mini-batch SGD iterations against the cost of one outer CA-SGD iteration.
CA-SGD trades $s$ small AllReduces for one grouped AllReduce of the pair $(r, G)$. This latency improvement comes at the cost of $(sb)^2 n_{\rm loc}$ additional Gram flops and an $(sb)^2$-word increase in the AllReduce message size.

\begin{table}[ht]
\centering
\caption{$\alpha$-$\beta$-$\gamma$ costs of $s$ vanilla SGD
iterations versus one outer CA-SGD iteration under 1D block-column
partitioning with AllReduce cost~\eqref{eq:allreduce-cost}.}
\label{tab:abg-comparison}
\scriptsize
\setlength{\tabcolsep}{3pt}
\begin{tabular}{@{}lll@{}}
\toprule
Cost component & SGD ($s$ iters) & CA-SGD (one outer iter) \\
\midrule
Flops $F$   & $4 s b\, n_{\text{loc}}$
            & $2 (sb)^{2} n_{\text{loc}} + b^2s(s-1) + 4 s b\, n_{\text{loc}}$ \\
Words $W$   & $s b$
            & $s b + (sb)^{2}$ \\
Messages $L$ & $s$
            & $1$ \\
\bottomrule
\end{tabular}
\end{table}

The two costs have a crossover at $\alpha(s{-}1)\cdot 2\log P =\gamma\bigl[2(sb)^2 n_{\rm loc}+b^2 s(s{-}1)\bigr] +\beta(sb)^2\cdot 2(P{-}1)/P$.
Mixed precision shrinks the right-hand-side Gram terms by decreasing message size and by increasing computation throughput through the use of tensor cores.
The inner-correction term $\gamma b^2s(s-1)$ stays unchanged because we keep this kernel at FP32 throughout (\Cref{tab:recipe}) as specified by the analysis.
Therefore, in the case where the Gram terms dominate the right-hand side, the crossover moves to smaller $s$ under mixed precision.

\section{Finite-Precision Analysis}
\label{sec:analysis}
This section presents the finite-precision error analysis of CA-SGD under the model described in \Cref{sec:notation}.
First, we bound the local forward error of a single outer CA-SGD iteration for the nine precision choices specified in \Cref{subsec:notation-nine-u}. Uniform precision and Recipe~C then appear as special cases of this bound.
Finally, a perturbed-SGD argument raises the per-block error budget to a convergence guarantee---convergence to a neighborhood of the optimum, valid as long as the iterates remain within the bounded region $\mathcal{X}_R$.
Our analysis covers the family of \emph{Lipschitz-residual GLMs}, which includes logistic regression, linear regression, and any member whose inverse-link derivative is globally Lipschitz.
Poisson regression lies outside this family because its inverse-link derivative is unbounded (\Cref{ass:lipschitz}). 
In addition, we include Poisson regression as empirical boundary-stress evidence beyond the loss family covered by the theorem.

\subsection{Assumptions}
\label{subsec:assumptions}

\begin{assumption}[Data normalization]
\label{ass:norm}
Every row of $A$ satisfies $\norm{a_i}_2\le 1$, which yields
$\fnorm{Y}\le\sqrt{sb}$ and $\norm{Y}_2\le\sqrt{sb}$ for every
sampled block $Y\in\R^{sb\times n}$.
\end{assumption}

\begin{assumption}[Lipschitz and bounded residual]
\label{ass:lipschitz}
Let $\mathcal X_R=\{x:\norm{x}_2\le R\}$ and
$\mathcal M_R=\{a_i^\top x:i\le m,x\in\mathcal X_R\}$.
The residual $\delta(r,y)=-\partial L/\partial r$ is separable
componentwise and globally $L_\sigma$-Lipschitz in $r$.
For the bounded-margin regime used to instantiate the local error
budget, the residual also satisfies $\abs{\delta}\le D_\delta$ and
$\abs{\mathrm{fl}_u(\delta)-\delta}\le C_\delta u D_\delta$ on the
exact and finite-precision margins encountered in the local comparison.
Logistic and linear regression satisfy this with
$(L_\sigma,D_\delta)=(1/4,1)$ and $(1,\max\abs{y_i}+R)$
respectively.
\end{assumption}

\begin{assumption}[Bounded Gram]
\label{ass:gram-bound}
$\fnorm{G}\le sb$ and $\fnorm{G_L}\le sb$ for every sampled
$G=YY^\top$ and its lower-triangular part $G_L$ (deterministic under
\Cref{ass:norm}).
\end{assumption}

\begin{figure}[ht]
\centering
\begin{tikzpicture}
\begin{semilogxaxis}[validation,
    xlabel={$P$ (number of ranks)},
    ylabel={$C(P)$ observed (max, $L\ge131072$)},
    legend pos=north west,
    xtick={4,8,16,32,64,128,256},
    xticklabels={4,8,16,32,64,128,256},
    ymin=0,
]
  \addplot[ColA, mark=*, thick] coordinates {(4,7.258) (8,19.77) (16,50.18) (32,73.09) (64,141.7) (128,326.3) (256,611.6) };
  \addlegendentry{FP32}
  \addplot[BinCA, mark=*, thick] coordinates {(4,7.301) (8,19.8) (16,52.2) (32,71.97) (64,154.7) (128,305) (256,588.6) };
  \addlegendentry{BF16}
  \addplot[BinSGD, mark=*, thick] coordinates {(4,7.433) (8,20.15) (16,52.5) (32,69.77) (64,154.6) (128,300.5) (256,610) };
  \addlegendentry{FP16}
  \addplot[gray, dotted, thick] coordinates {(4,10.00) (8,14.71) (16,21.64) (32,31.84) (64,46.85) (128,68.92) (256,101.40) };
  \addlegendentry{$4.62\,P^{0.56}$ (empirical fit, signed)}
  \addplot[black, dashed, thick] coordinates {(4,3) (8,4) (16,5) (32,6) (64,7) (128,8) (256,9) };
  \addlegendentry{$3+\log_2(P/4)$ (prior assumption)}
\end{semilogxaxis}
\end{tikzpicture}
\caption{AllReduce constant $C(P)$ measured directly on Perlmutter NCCL collectives versus rank count $P$, summing rank-local vectors of length $L\ge 131072$ against an FP64-accumulated reference. The observed factor reaches the worst-case $C(P)\lesssim 2.4P$ and is datatype-independent to within $4\%$ across BF16, FP16, and FP32. The dotted curve is the power-law fit $4.62P^{0.56}$ and the dashed curve is the prior $3+\log_2(P/4)$ assumption.}
\label{fig:nccl-cp}
\end{figure}
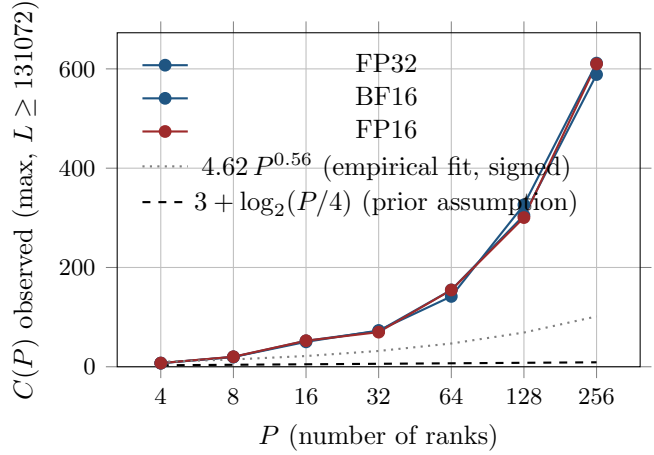

\begin{assumption}[AllReduce model]
\label{ass:allreduce}
For any AllReduce summing rank-local vectors $v=\sum_p v_p$ into
collective unit roundoff $u$, the returned $\widehat v$ satisfies
$\abs{\widehat v - v}\le C(P)\,u\,\sum_p\abs{v_p}+\bigO(u^2)$.
The constant $C(P)$ depends on processor count, topology, and the
sign structure of the reduction.
$C(P)$ absorbs the in-network reduction rounding and any input-cast
rounding that occurs when the rank-local accumulator is finer than
the collective datatype.
The bounds below hold for any finite $C(P)$.
When evaluating constants numerically we use the worst-case trace bound $C(P)\lesssim 2.4P$. The direct NCCL measurements of \Cref{fig:nccl-cp} reach this worst case ($C(256)\approx 6\!\times\!10^{2}$), while their empirical power-law fit $4.62P^{0.56}$ grows more slowly (to $\approx 10^{2}$ at $P=256$), and both are datatype-independent to within $4\%$.
\end{assumption}

\begin{assumption}[Bounded-trajectory regime]
\label{ass:weight}
All exact and finite-precision iterates and exact mini-batch substeps
considered in the local and global comparisons remain in $\mathcal X_R$.
\end{assumption}

\begin{assumption}[Optimization model]
\label{ass:global-opt}
$F(x)=m^{-1}\sum_i L(a_i^\top x,y_i)$ is $\mu_F$-strongly convex and
$L_F$-smooth on $\mathcal X_R$, with unique unconstrained minimizer
$x_\star=\arg\min_{x\in\R^n}F(x)$ lying in the interior of $\mathcal X_R$.
In particular $\nabla F(x_\star)=0$.
Mini-batches are i.i.d.\ uniform. For $x\in\mathcal X_R$, the stochastic
gradient $g_I(x)$ is unbiased with
$\E\norm{g_I(x)-\nabla F(x)}_2^2\le\sigma_g^2/b$.
The exact CA-SGD block with parameter $s$ is the composition of
$s$ mini-batch SGD steps with $\eta\le 1/L_F$.
\end{assumption}

\subsection{Local error bound}
\label{subsec:uniform}
\label{subsec:mixed}

\paragraph{Source quantities and amplification factor.}

First, we fix the notation used in the local bound. 
Let $\alpha = \eta / b$ denote the effective step size, and let $d = (d_0;\dots;d_{s-1}) \in \R^{sb}$ be the stacked exact residual history across the $s$ inner steps.
Let $G_L$ denote the strict block-lower triangular part of $G = YY^\top$.
$G_L$ is nilpotent within the block (its $s$-th power vanishes), so the matrix $I - L_\sigma \alpha \abs{G_L}$ has a finite Neumann inverse given by a terminating power series. We define the amplification factor as:
\begin{equation}
\label{eq:triangular-amplification}
 K_s := \norm{(I - L_\sigma \alpha \abs{G_L})^{-1}}_2.
\end{equation}
In addition, we need a non-negative block-lower matrix that replaces the Gram entries with their magnitudes to prevent sign cancellation.
Thus, we define $H_L$ as the strict block-lower matrix with $H_{j,i} = \abs{Y_j} \abs{Y_i}^\top$ for $i < j$.
Finally, we collect the four data-dependent source quantities that each precision slot multiplies:
\begin{equation}
\label{eq:source-quantities}
  T_g := \norm{\abs{Y}^{\!\top}\!\abs{d}}_2,\;
  T_r := \norm{\abs{Y}\abs{x}}_2,\;
  T_G := \norm{H_L\abs{d}}_2,\;
  T_c := \!\Bigl(\sum_{j=1}^{s-1}\!\bigl(jb\fnorm{G_{j,0:j-1}}\norm{d_{0:j-1}}_2\bigr)^{\!2}\Bigr)^{\!1/2}\!\!.
\end{equation}

\begin{theorem}[Mixed-precision local error]
\label{thm:mixed-precision}
Let each operation in the outer iteration carry its own unit
roundoff $u_*\in\{u_A,u_G,u_r,u_c,u_\sigma,u_g,
u_{\mathrm{AR},r},u_{\mathrm{AR},G},u_x\}$, with raw or effective
inner-product Higham factors $\gamma_r,\gamma_G,\gamma_g$ for the
margin, Gram, and gradient kernels (single-format kernels use
$\gamma_q^{(u_*)}$. Low-precision-input/FP32-accumulate kernels use
the analogous input-cast plus FP32-accumulation factor, with the BF16
case given in \Cref{eq:bf16-fp32-accum}).
Assume all deterministic Higham factors appearing below are in their
validity domain.
Under
\Cref{ass:norm,ass:lipschitz,ass:gram-bound,ass:allreduce,ass:weight}
for the local one-step comparison in which the exact and
finite-precision recurrences start from the same
$x\in\mathcal X_R$ and use the same sampled rows and labels, with
$\omega_*=\max\{u_A,\gamma_r,\gamma_G,\gamma_{(s-1)b}^{(u_c)},
\gamma_g,C(P)u_{\mathrm{AR},r},C(P)u_{\mathrm{AR},G},u_\sigma,u_x\}$
and $\tau_{G,*}=K_s L_\sigma\alpha
(2u_A+\gamma_G+C(P)u_{\mathrm{AR},G})\norm{H_L}_2$ both at most $1/2$,
the local forward error admits the first-order additive
decomposition
\begin{equation}
\label{eq:mixed-bound}
  \norm{\iterhat{s} - x^{(s)}}_2
  \;\le\; \sum_* E_* + \mathcal R_{\rm mix}^{(2)},
\end{equation}
with the nine leading terms
\begin{equation}
\label{eq:mixed-Es}
\begin{aligned}
  E_{u_{A}}        &= \alpha\norm{Y}_2 K_s L_\sigma u_A T_r
                    + 2\alpha^2\norm{Y}_2 K_s L_\sigma u_A T_G
                    + \alpha u_A T_g,\\
  E_{u_{r}}        &= \alpha\norm{Y}_2 K_s L_\sigma \gamma_r T_r,\\
  E_{u_{\mathrm{AR},r}}
                   &= \alpha\norm{Y}_2 K_s L_\sigma C(P) u_{\mathrm{AR},r} T_r,\\
  E_{u_{G}}        &= \alpha^2\norm{Y}_2 K_s L_\sigma \gamma_G T_G,\\
  E_{u_{\mathrm{AR},G}}
                   &= \alpha^2\norm{Y}_2 K_s L_\sigma C(P) u_{\mathrm{AR},G} T_G,\\
  E_{u_{c}}        &= \alpha\norm{Y}_2 K_s L_\sigma u_c T_r
                    + 2\alpha^2\norm{Y}_2 K_s L_\sigma u_c T_G
                    + \alpha^2\norm{Y}_2 K_s L_\sigma u_c T_c,\\
  E_{u_{\sigma}}   &= \alpha\norm{Y}_2 K_s C_\delta u_\sigma D_\delta\sqrt{sb},\\
  E_{u_{g}}        &= \alpha\gamma_g T_g,\\
  E_{u_{x}}        &= u_x\bigl(\norm{x^{(s)}}_2+\alpha\norm{Y^\top d}_2\bigr),
\end{aligned}
\end{equation}
and remainder
$\mathcal R_{\rm mix}^{(2)}\le C_{\rm loc}\Omega_*\sum_* E_*$ with
$\Omega_*=\max\{\omega_*,\tau_{G,*}\}$.
Per \Cref{ass:allreduce}, the constant $C(P)$ absorbs any input-cast
rounding that occurs when a recipe casts its rank-local accumulator
down to a coarser collective datatype.
\end{theorem}

\begin{corollary}[Uniform-precision local error]
\label{thm:uniform-precision}
Set all nine slots in \Cref{thm:mixed-precision} to the same unit
roundoff $u$, and take $\gamma_r=\gamma_G=\gamma_{n_{\rm loc}}^{(u)}$
and $\gamma_g=\gamma_{sb}^{(u)}$.
Then
\begin{equation}
\label{eq:uniform-bound}
  \norm{\iterhat{s}-x^{(s)}}_2
  \le E_{\rm step}+E_r+E_G+E_{\rm inner}+E_\sigma
  +\mathcal R_{\rm unif}^{(2)},
\end{equation}
where the five terms are obtained by grouping the nine rows of
\Cref{eq:mixed-Es} by operation family, and
$\mathcal R_{\rm unif}^{(2)}
\le C_{\rm loc}\Omega(u)(E_{\rm step}+E_r+E_G+E_{\rm inner}
+E_\sigma)$ with
$\Omega(u)=\max\{\omega(u),\tau_G(u)\}$,
$\omega(u)=\max\{\gamma_{n_{\rm loc}}^{(u)},\gamma_{sb}^{(u)},
\gamma_{(s-1)b}^{(u)},C(P)u,u\}$, and
$\tau_G(u)=K_sL_\sigma\alpha
(\gamma_{n_{\rm loc}}^{(u)}+(C(P)+2)u)\norm{H_L}_2$.
\end{corollary}

In Recipe~C, the Gram GEMM and the margin and gradient GEMVs use BF16 inputs with an FP32 accumulator on tensor cores.
This kernel has effective Higham factor
\begin{equation}
\label{eq:bf16-fp32-accum}
  \widetilde\gamma_n^{(u_{r,G,g})}
  := 2u_{16} + \gamma_n^{(u_{32})}
  \approx 2u_{16} + n u_{32},
\end{equation}
where $2u_{16}$ charges the two BF16 input conversions and $\gamma_n^{(u_{32})}$ charges the FP32 accumulator.
The experiments use reduction lengths $n_{\rm loc}\le 10^4$ and $sb\le 16384$ (\Cref{sec:experiments}), at which $nu_{32}\le 10^{-3}$.
Therefore the $2u_{16}\approx 8\!\times\!10^{-3}$ term dominates, and the BF16 GEMM/GEMV rows reduce to first order in the input conversion.

The correction row $E_{u_c}$ is the only row that mixes three sources.
The $T_c$ term carries the $jb$ factor from the correction GEMV.
Its length-$(s{-}1)b$ deterministic Higham tail is absorbed into $\mathcal R^{(2)}$, since $\gamma_{(s{-}1)b}^{(u_c)}\le\Omega_*$.
The $T_r$ and $T_G$ summands enter from the AXPY and the scalar-multiply roundings inside the correction step.
The hypothesis $(s{-}1)b\,u_c<1$ ensures that the deterministic factor $\gamma_{(s{-}1)b}^{(u_c)}$ multiplying $T_c$ is valid.
Our experiments are in the regime where $(s{-}1)b\,u_{16}>1$, so $u_c$ is set to FP32. The same FP32 choice keeps the $T_r$ and $T_G$ summands at first order.

\subsection{Global convergence}
\label{subsec:global}

\begin{theorem}[SGD convergence]
\label{thm:global-perturbed}
Under \Cref{ass:norm,ass:lipschitz,ass:weight,ass:global-opt}, let
$\Phi_h$ be the exact CA-SGD block map for outer block $h$ and let
the finite-precision block satisfy
$\widehat x_{h+1}=\Phi_h(\widehat x_h)+e_h$ with
$\norm{e_h}_2\le\rho$ deterministically, where $\rho$ is any uniform
upper bound on the local budget
$\sum_* E_*+\mathcal R_{\rm mix}^{(2)}$ from
\Cref{thm:mixed-precision}.
For a run of length $H=Ms$, set $a=(1-\mu_F\eta)^s$ and
$\nu_s=\eta\sigma_g^2(1-a)/(\mu_F b)$.
For any $\chi>0$ with $q=(1+\chi)a<1$,
\begin{equation}
\label{eq:global-distance}
  \E\norm{\widehat x_M-x_\star}_2^2
  \le q^M\norm{\widehat x_0-x_\star}_2^2
  +\frac{1-q^M}{1-q}\bigl((1+\chi)\nu_s
  +(1+\chi^{-1})\rho^2\bigr),
\end{equation}
and $\E[F(\widehat x_M)-F(x_\star)]
\le(L_F/2)\E\norm{\widehat x_M-x_\star}_2^2$.
Because \Cref{ass:weight} fixes $\mathcal X_R$ a posteriori, from the smallest ball containing the observed iterates rather than through an enforced projection, \Cref{thm:global-perturbed} certifies the reported runs rather than predicting boundedness.
Subject to that conditioning, a precision recipe that keeps the local block budget $\rho$ small inherits the exact mini-batch SGD rate up to a variance/rounding neighborhood while the bounded-trajectory and local-curvature hypotheses hold.
\end{theorem}

\paragraph{Recipe~C neighborhood.}
We specialize \Cref{thm:global-perturbed} to Recipe~C (\Cref{tab:recipe}).
The radius becomes $\rho_C=(1+C_{\rm loc}\Omega_C)S_C$, where $S_C$ is the sum of the nine leading terms of \Cref{eq:mixed-Es} under the Recipe~C precision assignment.
On logistic regression with the parameters fixed to $(b,s,P,n_{\rm loc},\eta)=(32,64,16,256,0.5)$, we obtain $\Omega_C\le 1.6\!\times\!10^{-1}$, with the bound $\rho_C^{\rm det}=(1+C_{\rm loc}\Omega_C)S_C$ finite when $\Omega_C\le 1/2$.

The dominant contribution to $\Omega_C$ is the BF16 margin-AllReduce term $C(P)\,u_{\mathrm{AR},r}$, which grows with $P$, and this validation fixes $P=16$, where $C(16)\,u_{16}\approx 1.5\!\times\!10^{-1}$.
The measured growth $C(P)\approx 4.62P^{0.56}$ (\Cref{fig:nccl-cp}) keeps $C(P)\,u_{16}<1/2$ through $P=256$, so the $\Omega_C\le 1/2$ hypothesis of \Cref{thm:mixed-precision} holds at every scale we run. The worst-case bound $C(P)\lesssim 2.4P$ reaches the threshold near $P\approx 60$ and the measured fit near $P\approx 330$.
Beyond that scale Recipe~C must promote the margin AllReduce $u_{\mathrm{AR},r}$ to FP32 (the $f$ column of \Cref{tab:recipe-configs}) to keep the hypothesis. A direct validation of the local budget at $P=256$, where the accuracy comparison of \Cref{tab:expt-thm3-validation} is not run, is left for future work.

\begin{table}[ht]
\centering
\caption{Precision symbols for the nine slots used in \Cref{sec:experiments}.
For storage, collectives, and non-GEMM/GEMV rows, the symbols denote unit roundoffs.
In the $u_G$, $u_r$, and $u_g$ rows, $b$ denotes BF16 input with FP32 accumulation, $h$ denotes FP16 input with FP32 accumulation, and $h_{\rm a}$ denotes FP16 input/output with FP16 accumulation.
These kernel rows are evaluated through the corresponding Higham factors in \Cref{thm:mixed-precision}. The factor $h_{\rm a}$ uses $\gamma_q^{(u_{\rm FP16})}$ rather than $2u_{\rm FP16}+\gamma_q^{(u_{\rm FP32})}$.
H and I are low-precision only baselines.
Symbols:
$f=u_{\rm FP32}\approx 6\!\times\!10^{-8}$,
$t=u_{\rm TF32}\approx 5\!\times\!10^{-4}$,
$h=u_{\rm FP16}\approx 5\!\times\!10^{-4}$,
$b=u_{\rm BF16}\approx 4\!\times\!10^{-3}$.}
\label{tab:recipe-configs}
\label{tab:recipe}
\small
\begin{tabular}{@{}lccccccccc@{}}
\toprule
$u_{\ast}$ & A & B & C & D & E & F & G & H & I \\
\midrule
$u_A$            & $f$ & $b$ & $b$ & $b$ & $h$ & $h$ & $f$ & $h$ & $b$ \\
$u_G$            & $f$ & $f$ & $b$ & $b$ & $h$ & $h_{\rm a}$ & $t$ & $h_{\rm a}$ & $b$ \\
$u_r$            & $f$ & $f$ & $b$ & $b$ & $h$ & $h_{\rm a}$ & $t$ & $h_{\rm a}$ & $b$ \\
$u_c$            & $f$ & $f$ & $f$ & $f$ & $f$ & $f$ & $t$ & $h$ & $b$ \\
$u_\sigma$       & $f$ & $f$ & $f$ & $f$ & $f$ & $f$ & $f$ & $h$ & $b$ \\
$u_g$            & $f$ & $f$ & $b$ & $b$ & $h$ & $h_{\rm a}$ & $t$ & $h_{\rm a}$ & $b$ \\
$u_{\mathrm{AR},r}$ & $f$ & $f$ & $b$ & $b$ & $h$ & $h$ & $f$ & $h$ & $b$ \\
$u_{\mathrm{AR},G}$ & $f$ & $f$ & $f$ & $b$ & $f$ & $f$ & $f$ & $h$ & $b$ \\
$u_x$            & $f$ & $f$ & $f$ & $f$ & $f$ & $f$ & $f$ & $h$ & $b$ \\
\bottomrule
\end{tabular}
\end{table}

\section{Implementation on NVIDIA A100}
\label{sec:implementation}

The kernel DAG of \Cref{fig:dag-casgd-mp} and the pseudocode of \Cref{alg:mp-ca-sgd} fix the algorithmic structure at the kernel level.
This section describes the concrete CUDA~\cite{cuda_doc}, cuBLAS~\cite{cublas_doc}, and NCCL~\cite{nccl_doc} choices that realize the mixed-precision implementation on NVIDIA A100 benchmarked in \Cref{sec:experiments}.

We run every benchmark of \Cref{sec:experiments} on the GPU partition of NERSC Perlmutter, an HPE Cray EX system.
Each Perlmutter GPU node pairs a single AMD EPYC 7763 (Milan, 64 cores, 256\,GB DDR4) host with four NVIDIA A100 SXM4-40GB devices~\cite{nvidia2021a100}.
Each A100 exposes 108 SMs, 40\,GB of HBM2e at 2039\,GB/s, and the FP32, TF32, and BF16 compute roofs of \Cref{sec:perf-model}.
Every device runs at the default 400\,W power cap throughout.
Within a node, the four GPUs form a fully-connected NVLink~3 mesh with twelve $25$\,GB/s links per device, giving $600$\,GB/s aggregate peer-to-peer bidirectional bandwidth.
This mesh keeps the intra-node AllReduce of \Cref{sec:comm-cost} on NVLink and off the PCIe path.
Across nodes, communication runs over HPE Slingshot~11, with four $200$\,Gbps Cassini NICs per GPU node on a three-hop dragonfly fabric.

The implementation applies no explicit projection or norm clip on $\widetilde x$.
The local budgets are instantiated with the radius $R$ of \Cref{ass:weight}, taken a posteriori as the smallest ball containing the observed iterates of the reported run.
\Cref{ass:weight} held on every run reported here, so the convergence theorem is a conditional guarantee that certifies these runs rather than predicting boundedness. Were an iterate to leave $\mathcal X_R$, the local-curvature constants $\mu_F,L_F$ of \Cref{ass:global-opt} and hence the radius $\rho$ would no longer be controlled, and a projection step would be required to restore the hypothesis.
We fuse the outer gradient GEMV and the weight update of \Cref{alg:mp-ca-sgd} (lines \ref{line:gradient}--\ref{line:update}) into a single BF16-input/FP32-accumulate GEMM with $\alpha=\eta/b$ and $\beta=1$.
This GEMM adds the gradient directly into the FP32 master weights.
The fusion leaves the precision-slot accounting of \Cref{thm:mixed-precision} unchanged.

Every BF16-input GEMM and GEMV uses FP32 accumulation.
The Gram GEMM avoids cuBLAS's TF32 ``fast-FP32'' fallback so that the $s^2$ off-diagonal entries retain true FP32 precision.

We issue a single grouped AllReduce that carries both $r$ and $G$ at their distinct datatypes (BF16 and FP32 in Recipe~C).
This collective costs one $\alpha\cdot 2\log p$ latency per outer iteration.
Compared to an unfused pair of AllReduces, the grouped form halves the latency cost (\Cref{sec:comm-cost}).

Each optimizer records two CUDA-graph executables that bracket the collective.
The pre-collective graph runs sampling, gather, and the local cuBLAS kernels.
The post-collective graph runs the residual, correction, and weight update.
The grouped AllReduce is issued between the two graph launches.

\section{Numerical Experiments}
\label{sec:experiments}

\providecommand{\plotsdir}{../experiments/plots}
\providecommand{\resultsdir}{../experiments/results}
\IfFileExists{experiments/plots/pgfplots-style.tex}{
\pgfplotsset{compat=1.18}

\definecolor{ColA}{HTML}{1F5582}
\definecolor{ColC}{HTML}{9E2A2B}

\definecolor{BinSGD}{HTML}{9E2A2B}
\definecolor{BinSGDBF}{HTML}{E89090}
\definecolor{BinCA}{HTML}{1F5582}
\definecolor{BinCATF}{HTML}{3E7CB1}
\definecolor{BinCABF}{HTML}{7DB7D9}
\definecolor{BinCABFD}{HTML}{ABC7DA}
\definecolor{BinCAFP16}{HTML}{5B9E9E}

\pgfplotsset{
  validation/.style={
    width=8.5cm, height=6cm,
    grid=major, tick align=outside,
    every axis legend/.append style={font=\small, draw=none, fill=none},
    label style={font=\small},
    tick label style={font=\footnotesize},
    title style={font=\small},
  }
}

\pgfplotsset{
  scaling/.style={
    width=8.5cm, height=6cm,
    grid=major, tick align=outside,
    every axis legend/.append style={font=\small, draw=none, fill=none},
    label style={font=\small},
    tick label style={font=\footnotesize},
    title style={font=\small},
    cycle list={
      {BinSGD,    mark=*,         thick, mark size=2pt},
      {BinSGDBF,  mark=triangle*, thick, mark size=2.5pt},
      {BinCA,     mark=square*,   thick, mark size=2pt},
      {BinCATF,   mark=square,    thick, mark size=2pt,   dashed},
      {BinCABF,   mark=diamond*,  thick, mark size=2.5pt},
      {BinCABFD,  mark=diamond,   thick, mark size=2.5pt, dashed},
      {BinCAFP16, mark=asterisk,  thick, mark size=2.5pt, dotted},
    },
  }
}
}{%
  \definecolor{ColA}{HTML}{1F5582}\definecolor{ColC}{HTML}{9E2A2B}
  \pgfplotsset{validation/.style={width=8.5cm,height=6cm,grid=major}}}
\providecommand{\exptASlope}{0.465}
\providecommand{\thmOneCoherenceNloc}{625}

\subsection{Setup}
\label{subsec:platform}

We partition the feature matrix 1D column-block with $n_{\rm loc}=n/P$.
\Cref{tab:datasets} lists the benchmarks.
We $\ell_2$-normalize every row to satisfy \Cref{ass:norm}.
For each dataset we lock the step size $\eta^\star$ from a FP32 SGD sweep at $(b,s)=(32,1)$, and we reuse this $\eta^\star$ across every CA-SGD timing.
Unless noted otherwise, every run uses three seeds (42, 43, 44).

\begin{table}[ht]
\centering
\caption{Benchmark datasets. \texttt{synth} and
\texttt{Poisson-synth} share the same generator and feature matrix.}
\label{tab:datasets}
\small
\begin{tabular}{l r r l l}
\toprule
Dataset & $m$ & $n$ & Loss & Source \\
\midrule
\texttt{synth}        & $10^6$         & $9{,}984$            & logistic & synthetic \\
\texttt{Poisson-synth}     & $10^6$         & $9{,}984$            & Poisson  & synthetic \\
\texttt{epsilon}           & $4{\times}10^5$ & $2{,}000$           & logistic & LIBSVM \\
\texttt{SUSY}              & $5{\times}10^6$ & $18$                & logistic & LIBSVM \\
\texttt{HIGGS}             & $1.1{\times}10^7$ & $28$              & logistic & LIBSVM \\
\bottomrule
\end{tabular}
\end{table}

\subsection{Accuracy gap}
\label{subsec:accuracy-gap}

We compare Recipe~C against Recipe~A under a $0.5\%$ empirical validation threshold on the final loss.
The validation cells use $(m,n)=(65{,}536,4096)$, $b=32$, $s\in\{16,64\}$, $H=200$ outer iterations, and $P=16$.
\Cref{tab:expt-thm3-validation} reports the results.
The Lipschitz-residual logistic and linear cells covered by the local finite-precision theorem pass the threshold. The largest listed gap is $1.1\!\times\!10^{-4}$.
The Poisson rows also pass, but they sit outside \Cref{ass:lipschitz} and are reported only as boundary-stress evidence.

\begin{table}[ht]
\centering
\caption{Recipe~C verification on relative final-loss difference $|L_C-L_A|/|L_A|$ at $P=16$,
$(m,n)=(65{,}536,4096)$, $b=32$, $H=200$, three seeds.
$^\dagger$Poisson does not satisfy \Cref{ass:lipschitz}.}
\label{tab:expt-thm3-validation}
\small
\begin{tabular}{l r r}
\toprule
Loss & $s=16$ & $s=64$ \\
\midrule
logistic          & $0$                    & $0$ \\
linear            & $7\!\times\!10^{-5}$   & $1.1\!\times\!10^{-4}$ \\
Poisson$^\dagger$ & $1.3\!\times\!10^{-4}$ & $3.3\!\times\!10^{-4}$ \\
\bottomrule
\end{tabular}
\end{table}

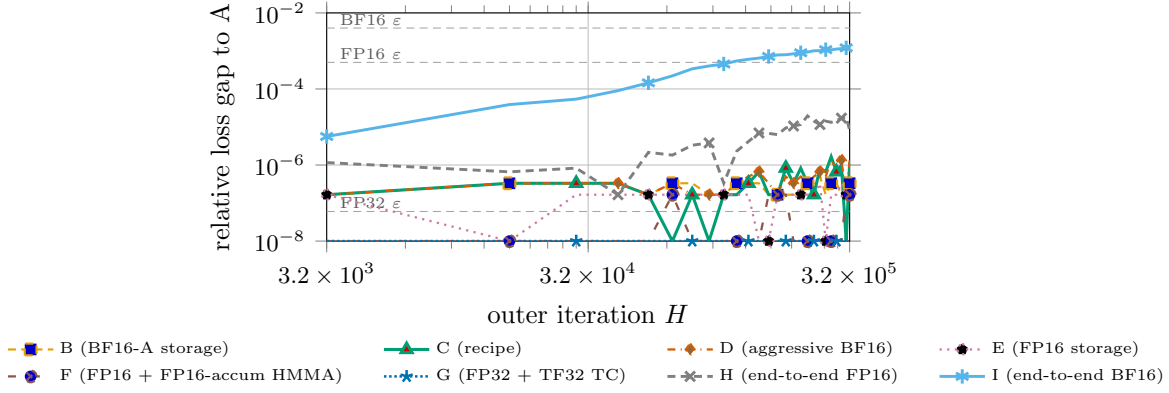
\begin{figure}[ht]
\centering
\definecolor{cfgA}{HTML}{000000}%
\definecolor{cfgB}{HTML}{E69F00}%
\definecolor{cfgC}{HTML}{009E73}%
\definecolor{cfgD}{HTML}{D55E00}%
\definecolor{cfgE}{HTML}{CC79A7}%
\definecolor{cfgF}{HTML}{8C564B}%
\definecolor{cfgG}{HTML}{0072B2}%
\definecolor{cfgH}{HTML}{7F7F7F}%
\definecolor{cfgI}{HTML}{56B4E9}%
\begin{tikzpicture}
\begin{axis}[
  validation,
  width=8.5cm, height=4.6cm,
  xmode=log, xmin=3200, xmax=320000,
  xtick={3200,32000,320000},
  xticklabels={$3.2\times10^3$,$3.2\times10^4$,$3.2\times10^5$},
  legend cell align=left,
  xlabel={outer iteration $H$},
  ylabel={relative loss gap to A},
  ymode=log, ymin=1e-8, ymax=1e-2,
  ytick={1e-8,1e-6,1e-4,1e-2},
  legend to name=exptbgap-legend,
  legend columns=4,
  legend style={font=\tiny, /tikz/every even column/.append style={column sep=0.45cm}},
]
\addplot[gray!70, densely dashed, line width=0.4pt, forget plot]
  coordinates {(3200,6e-8) (320000,6e-8)}
  node[pos=0.02, anchor=south west, font=\tiny, gray!70!black, inner sep=1pt] {FP32 $\varepsilon$};
\addplot[gray!70, densely dashed, line width=0.4pt, forget plot]
  coordinates {(3200,5e-4) (320000,5e-4)}
  node[pos=0.02, anchor=south west, font=\tiny, gray!70!black, inner sep=1pt] {FP16 $\varepsilon$};
\addplot[gray!70, densely dashed, line width=0.4pt, forget plot]
  coordinates {(3200,4e-3) (320000,4e-3)}
  node[pos=0.02, anchor=south west, font=\tiny, gray!70!black, inner sep=1pt] {BF16 $\varepsilon$};
\addplot+[draw=cfgB, mark=square*, densely dashed, mark size=2.2pt, line width=0.8pt, mark repeat=4, mark phase=2]
  coordinates {
    (3200,1.654910e-07)
    (16000,3.310030e-07)
    (28800,3.311072e-07)
    (41600,3.311894e-07)
    (54400,1.656139e-07)
    (67200,3.313074e-07)
    (80000,3.314067e-07)
    (92800,1.657219e-07)
    (105600,1.657366e-07)
    (118400,3.315506e-07)
    (131200,3.315953e-07)
    (144000,3.316276e-07)
    (156800,1.658349e-07)
    (169600,1.658570e-07)
    (182400,1.658596e-07)
    (195200,1.658641e-07)
    (208000,1.658872e-07)
    (220800,3.318068e-07)
    (233600,3.318104e-07)
    (246400,4.977138e-07)
    (259200,1.659256e-07)
    (272000,3.318451e-07)
    (284800,3.318902e-07)
    (297600,1.659387e-07)
    (310400,3.319407e-07)
    (319999,3.318848e-07)
  };
\addlegendentry{B (BF16-A storage)}
\addplot+[draw=cfgC, mark=triangle*, solid, mark size=2.4pt, line width=1.15pt, mark repeat=4, mark phase=3]
  coordinates {
    (3200,1.654910e-07)
    (16000,3.310030e-07)
    (28800,3.311072e-07)
    (41600,3.311894e-07)
    (54400,1.656139e-07)
    (67200,1.000000e-08)
    (80000,1.657034e-07)
    (92800,1.000000e-08)
    (105600,1.657366e-07)
    (118400,1.657753e-07)
    (131200,3.315953e-07)
    (144000,6.632551e-07)
    (156800,1.658349e-07)
    (169600,1.658570e-07)
    (182400,8.292978e-07)
    (195200,3.317282e-07)
    (208000,8.294358e-07)
    (220800,3.318068e-07)
    (233600,1.659052e-07)
    (246400,8.295231e-07)
    (259200,6.637026e-07)
    (272000,1.659225e-06)
    (284800,6.637803e-07)
    (297600,9.956322e-07)
    (310400,1.000000e-08)
    (319999,1.161597e-06)
  };
\addlegendentry{C (recipe)}
\addplot+[draw=cfgD, mark=diamond*, dashdotted, mark size=2.2pt, line width=0.85pt, mark repeat=4, mark phase=4]
  coordinates {
    (3200,1.654910e-07)
    (16000,3.310030e-07)
    (28800,3.311072e-07)
    (41600,3.311894e-07)
    (54400,1.656139e-07)
    (67200,3.313074e-07)
    (80000,1.657034e-07)
    (92800,1.657219e-07)
    (105600,1.657366e-07)
    (118400,3.315506e-07)
    (131200,4.973929e-07)
    (144000,6.632551e-07)
    (156800,3.316697e-07)
    (169600,1.658570e-07)
    (182400,4.975787e-07)
    (195200,3.317282e-07)
    (208000,4.976615e-07)
    (220800,3.318068e-07)
    (233600,4.977157e-07)
    (246400,6.636184e-07)
    (259200,4.977769e-07)
    (272000,1.327380e-06)
    (284800,8.297254e-07)
    (297600,1.327510e-06)
    (310400,1.659704e-07)
    (319999,1.327539e-06)
  };
\addlegendentry{D (aggressive BF16)}
\addplot+[draw=cfgE, mark=pentagon*, dotted, mark size=2.2pt, line width=0.85pt, mark repeat=4, mark phase=1]
  coordinates {
    (3200,1.654910e-07)
    (16000,1.000000e-08)
    (28800,1.655536e-07)
    (41600,1.655947e-07)
    (54400,1.656139e-07)
    (67200,1.656537e-07)
    (80000,1.657034e-07)
    (92800,1.657219e-07)
    (105600,1.657366e-07)
    (118400,1.657753e-07)
    (131200,1.657976e-07)
    (144000,1.000000e-08)
    (156800,1.000000e-08)
    (169600,1.658570e-07)
    (182400,1.658596e-07)
    (195200,1.658641e-07)
    (208000,1.658872e-07)
    (220800,1.659034e-07)
    (233600,1.659052e-07)
    (246400,3.318092e-07)
    (259200,1.000000e-08)
    (272000,1.659225e-07)
    (284800,1.659451e-07)
    (297600,3.318774e-07)
    (310400,1.659704e-07)
    (319999,3.318848e-07)
  };
\addlegendentry{E (FP16 storage)}
\addplot+[draw=cfgF, mark=otimes*, loosely dashed, mark size=2.2pt, line width=0.85pt, mark repeat=4, mark phase=2]
  coordinates {
    (3200,1.000000e-08)
    (16000,1.000000e-08)
    (28800,1.000000e-08)
    (41600,1.000000e-08)
    (54400,1.000000e-08)
    (67200,1.656537e-07)
    (80000,1.000000e-08)
    (92800,1.000000e-08)
    (105600,1.000000e-08)
    (118400,1.000000e-08)
    (131200,1.000000e-08)
    (144000,1.000000e-08)
    (156800,1.658349e-07)
    (169600,1.658570e-07)
    (182400,1.658596e-07)
    (195200,1.000000e-08)
    (208000,1.000000e-08)
    (220800,1.000000e-08)
    (233600,1.000000e-08)
    (246400,1.000000e-08)
    (259200,1.000000e-08)
    (272000,1.000000e-08)
    (284800,1.000000e-08)
    (297600,1.000000e-08)
    (310400,1.659704e-07)
    (319999,1.659424e-07)
  };
\addlegendentry{F (FP16 + FP16-accum HMMA)}
\addplot+[draw=cfgG, mark=star, densely dotted, mark size=2.4pt, line width=0.95pt, mark repeat=4, mark phase=3]
  coordinates {
    (3200,1.000000e-08)
    (16000,1.000000e-08)
    (28800,1.000000e-08)
    (41600,1.000000e-08)
    (54400,1.000000e-08)
    (67200,1.000000e-08)
    (80000,1.000000e-08)
    (92800,1.000000e-08)
    (105600,1.000000e-08)
    (118400,1.000000e-08)
    (131200,1.000000e-08)
    (144000,1.000000e-08)
    (156800,1.000000e-08)
    (169600,1.000000e-08)
    (182400,1.000000e-08)
    (195200,1.000000e-08)
    (208000,1.000000e-08)
    (220800,1.000000e-08)
    (233600,1.000000e-08)
    (246400,1.000000e-08)
    (259200,1.000000e-08)
    (272000,1.000000e-08)
    (284800,1.000000e-08)
    (297600,1.000000e-08)
    (310400,1.000000e-08)
    (319999,1.000000e-08)
  };
\addlegendentry{G (FP32 + TF32 TC)}
\addplot+[draw=cfgH, mark=x, densely dashed, mark size=2.6pt, line width=1.1pt, mark repeat=4, mark phase=4]
  coordinates {
    (3200,1.158437e-06)
    (16000,6.620060e-07)
    (28800,8.277680e-07)
    (41600,1.655947e-07)
    (54400,2.152981e-06)
    (67200,1.822191e-06)
    (80000,3.314067e-06)
    (92800,3.811603e-06)
    (105600,3.314732e-07)
    (118400,2.320854e-06)
    (131200,4.144941e-06)
    (144000,6.964179e-06)
    (156800,6.799230e-06)
    (169600,6.302566e-06)
    (182400,9.453995e-06)
    (195200,1.061530e-05)
    (208000,1.161210e-05)
    (220800,1.957660e-05)
    (233600,1.426785e-05)
    (246400,1.161332e-05)
    (259200,1.426961e-05)
    (272000,1.327380e-05)
    (284800,1.460317e-05)
    (297600,1.709169e-05)
    (310400,1.825674e-05)
    (319999,8.297121e-06)
  };
\addlegendentry{H (end-to-end FP16)}
\addplot+[draw=cfgI, mark=asterisk, solid, mark size=2.7pt, line width=1.1pt, mark repeat=4, mark phase=1]
  coordinates {
    (3200,5.626694e-06)
    (16000,3.889285e-05)
    (28800,5.397047e-05)
    (41600,9.024912e-05)
    (54400,1.450778e-04)
    (67200,2.211477e-04)
    (80000,3.388634e-04)
    (92800,4.023727e-04)
    (105600,4.541183e-04)
    (118400,5.455665e-04)
    (131200,6.031718e-04)
    (144000,6.488293e-04)
    (156800,7.159091e-04)
    (169600,7.851671e-04)
    (182400,7.952966e-04)
    (195200,8.410967e-04)
    (208000,8.947954e-04)
    (220800,9.333725e-04)
    (233600,1.010695e-03)
    (246400,1.016829e-03)
    (259200,1.081006e-03)
    (272000,1.066550e-03)
    (284800,1.120793e-03)
    (297600,1.151946e-03)
    (310400,1.206439e-03)
    (319999,1.187318e-03)
  };
\addlegendentry{I (end-to-end BF16)}
\end{axis}
\node[anchor=north, yshift=-1.05cm] at (current axis.south) {\pgfplotslegendfromname{exptbgap-legend}};
\end{tikzpicture}
\caption{Relative loss gap to Recipe~A on a long-horizon synthetic
logistic regression stress run with $P=4$, $b=32$, $s=16$,
$H=320{,}000$, and $\eta=0.10$. Values below $10^{-8}$ are plotted at $10^{-8}$.
Recipes~A--G follow the precision choices in \Cref{tab:recipe-configs}.
Recipes~H and~I intentionally perform end-to-end FP16 and BF16 training, respectively.}
\Description{Single-panel plot of the relative loss gap to the FP32 Recipe A on a log scale spanning 1e-8 to 1e-2 versus outer iteration H, overlaid with dashed horizontal reference lines at the FP32, FP16, and BF16 machine epsilons. Theorem-compliant Recipes A-G remain at or below the FP32 epsilon line, the end-to-end FP16 Recipe H sits between the FP16 and BF16 epsilon lines, and the end-to-end BF16 Recipe I sits below the BF16 epsilon line.}
\label{fig:expt-thm2-traj}
\end{figure}

\Cref{fig:expt-thm2-traj} extends the parity check to a long-horizon synthetic logistic regression stress run.
Recipes~A--G stay at the FP32 baseline to within $1.4\!\times\!10^{-6}$, which sits two orders of magnitude below the FP16 unit roundoff and over three orders below the BF16 unit roundoff.
The end-to-end Recipe~H lands at $8.3\!\times\!10^{-6}$ and Recipe~I at $1.19\!\times\!10^{-3}$, both above the mixed-precision recipes.

We repeat the accuracy check on mini-batch SGD run end to end in BF16 and FP16, measuring the worst final-loss gap to FP32 mini-batch SGD over three seeds and $s\in\{16,64\}$ at the operating point of \Cref{tab:expt-thm3-validation}.
The logistic gap is $7\!\times\!10^{-6}$ in both BF16 and FP16, the linear gap is $1.4\!\times\!10^{-4}$ (BF16) and $2.9\!\times\!10^{-4}$ (FP16), and the Poisson gap, outside \Cref{ass:lipschitz}, is $8.8\!\times\!10^{-4}$ (BF16) and $5.4\!\times\!10^{-4}$ (FP16). Every cell stays below the $0.5\%$ validation threshold.

\subsection{Speedup vs $(b,s)$ parameter grid}
\label{subsec:speedup}

We perform a sweep on $(b,s)$ to identify both the peak speedup and the working point we use for the real-data evaluation.
\Cref{tab:expt-d} reports the BF16 Recipe~C speedup over FP32 SGD on \texttt{synth}.
The fastest point in the grid is $(b,s)=(1,256)$ at $16.57\times$.
We use $(b,s)=(8,64)$ as the real-data operating point, which keeps $b=8$ for the smaller feature dimensions in \Cref{tab:datasets}. The locating sweep of \Cref{tab:expt-d} (three seeds, $H=200$) records $7.15\times$ on \texttt{synth} here, while the five-seed median end-to-end speedup for Recipe~C is $6.19\times$ (\Cref{tab:expt-e}).

\begin{table}[ht]
\centering
\caption{CA-SGD BF16 (Recipe~C) speedup over FP32 SGD (baseline: $b=32$,
$s=1$) on synth ($m=10^6$, $n=9{,}984$, $P=64$), averaged over three seeds
at $H=200$ to locate the operating point. The five-seed median end-to-end
speedups are in \Cref{tab:expt-e}. Best cell highlighted per
row. $b=128$ rows omitted.}
\label{tab:expt-d}
\small
\begin{tabular}{r r r r r r}
\toprule
$b$ & $s=1$ & $s=4$ & $s=16$ & $s=64$ & $s=256$ \\
\midrule
1   & $0.92\times$ & $3.13\times$ & $9.20\times$ & $14.62\times$ & $\mathbf{16.57\times}$ \\
8   & $0.87\times$ & $2.32\times$ & $4.95\times$ & $\mathbf{7.15\times}$ & $6.29\times$ \\
32  & $0.67\times$ & $1.46\times$ & $2.34\times$ & $\mathbf{2.29\times}$ & $1.28\times$ \\
\bottomrule
\end{tabular}
\end{table}

\subsection{Real-data speedup}
\label{subsec:expt-e}

We evaluate Recipes~A, G, C, D, and~F against FP32 SGD on five datasets at $(b,s)=(8,64)$ from \Cref{subsec:speedup}.
We report the median speedup over five random number generator seeds.
\Cref{tab:expt-e} reports the results.
Across these rows, the variants shown reach $5.1$--$6.8\times$ speedup over FP32 SGD. The theory-covered Recipe~C reaches $5.2$--$6.2\times$.
Recipe~D gives the largest speedup on \texttt{epsilon}, \texttt{synth}, and \texttt{Poisson-synth} and occupies the upper end of this range on the large-$n$ datasets, exceeding Recipe~C by at most $12\%$.
Recipe~D differs from Recipe~C only in the Gram AllReduce, which D casts to BF16 ($u_{\mathrm{AR},G}=b$ in \Cref{tab:recipe-configs}).
We recommend Recipe~C as the theory-covered default and Recipe~D as a faster large-$n$ variant whose Gram-collective error needs a per-problem check.

\begin{table}[ht]
\centering
\caption{Speedup at $(b,s)=(8,64)$ over FP32 SGD
($b=32,s=1$), median runtime is reported over five seeds.
Recipe~C is the theory-covered default.
$^\dagger$\texttt{Poisson-synth} does not satisfy \Cref{ass:lipschitz}.}
\label{tab:expt-e}
\small
\begin{tabular}{l l r r r r r r}
\toprule
Dataset & $P$ & SGD (ms) & A & G & C & D & F \\
\midrule
epsilon            & 16 & 0.055 & $5.40\times$ & $5.24\times$ & $5.56\times$ & $5.86\times$ & $5.56\times$ \\
SUSY               & 4  & 0.027 & $5.31\times$ & $5.31\times$ & $5.21\times$ & $5.31\times$ & $5.11\times$ \\
HIGGS              & 4  & 0.028 & $5.33\times$ & $5.33\times$ & $5.23\times$ & $5.23\times$ & $5.13\times$ \\
synth         & 64 & 0.085 & $5.85\times$ & $6.01\times$ & $6.19\times$ & $6.68\times$ & $6.24\times$ \\
Poisson-synth$^\dagger$ & 64 & 0.085 & $5.97\times$ & $6.01\times$ & $6.10\times$ & $6.84\times$ & $6.19\times$ \\
\bottomrule
\end{tabular}
\end{table}

\subsection{Strong and weak scaling}
\label{subsec:expt-scaling}

Strong scaling fixes the total problem at $(m,n)=(65{,}536,524{,}288)$ and partitions it across $P\in\{4,8,16,32,64,128,256\}$ GPUs, so $n_{\rm loc}$ shrinks from $131{,}072$ to $2{,}048$.
Weak scaling fixes the per-rank work at $(m, n_{\rm loc})=(65{,}536,65{,}536)$ and grows the total $n$ with $P$, from $262{,}144$ to $1.7\!\times\!10^{7}$.
Both experiments use synthetic logistic regression data with $\ell_2$-normalized rows (\Cref{ass:norm}), $b=32$, and $H=100$ outer iterations per run.
In \Cref{fig:scaling}, CA-SGD with BF16 storage reaches $0.062$~ms per equivalent SGD iteration at $P=128$.
At $P=256$, CA-SGD with BF16 storage reaches $0.741$ weak-scaling efficiency, compared with $0.668$ for FP32 SGD.
The BF16-storage-only mini-batch SGD variant tracks the FP32 mini-batch SGD baseline to within $0.9\%$, separating the scaling effect of the CA reformulation from the precision change alone.

\definecolor{scaleSgdFp32}{HTML}{666666}
\definecolor{scaleSgdBf16}{HTML}{E69F00}
\definecolor{scaleCfgA}{HTML}{000000}
\definecolor{scaleCfgC}{HTML}{009E73}
\definecolor{scaleCfgD}{HTML}{D55E00}
\definecolor{scaleCfgF}{HTML}{8C564B}
\definecolor{scaleCfgG}{HTML}{0072B2}

\tikzset{
    scalingIdeal/.style={
        black,
        dashed,
        line width=0.75pt,
        mark=none,
    },
    scalingSgdFp32/.style={
        draw=scaleSgdFp32,
        solid,
        mark=*,
        line width=0.8pt,
        mark size=2.4pt,
        mark options={solid, fill=scaleSgdFp32, draw=scaleSgdFp32},
    },
    scalingSgdBf16/.style={
        draw=scaleSgdBf16,
        loosely dotted,
        mark=square*,
        line width=0.85pt,
        mark size=2.6pt,
        mark options={solid, fill=white, draw=scaleSgdBf16, line width=0.65pt},
    },
    scalingCfgA/.style={
        draw=scaleCfgA,
        densely dashed,
        mark=triangle*,
        line width=0.95pt,
        mark size=2.8pt,
        mark options={solid, fill=white, draw=scaleCfgA, line width=0.65pt},
    },
    scalingCfgC/.style={
        draw=scaleCfgC,
        solid,
        mark=diamond*,
        line width=1.25pt,
        mark size=3.0pt,
        mark options={solid, fill=scaleCfgC, draw=scaleCfgC},
    },
    scalingCfgG/.style={
        draw=scaleCfgG,
        dash pattern=on 4pt off 1.4pt on 1pt off 1.4pt,
        mark=pentagon*,
        line width=1.0pt,
        mark size=2.9pt,
        mark options={solid, fill=white, draw=scaleCfgG, line width=0.65pt},
    },
    scalingCfgD/.style={
        draw=scaleCfgD,
        dashdotted,
        mark=star,
        line width=1.0pt,
        mark size=3.1pt,
        mark options={solid, fill=white, draw=scaleCfgD, line width=0.65pt},
    },
    scalingCfgF/.style={
        draw=scaleCfgF,
        loosely dashed,
        mark=otimes*,
        line width=1.0pt,
        mark size=2.9pt,
        mark options={solid, fill=white, draw=scaleCfgF, line width=0.65pt},
    },
}

\begin{figure}[ht]
\centering
\begin{subfigure}[t]{0.49\linewidth}
\centering
\begin{tikzpicture}
\begin{loglogaxis}[
    log basis x=2, log basis y=2,
    width=0.92\linewidth,
    height=0.62\linewidth,
    xlabel={GPUs ($P$)},
    ylabel={Runtime (ms/equiv-iter)},
    xmin=3, xmax=320,
    ymin=0.005, ymax=0.6,
    xtick={4,8,16,32,64,128,256},
    xticklabels={4,8,16,32,64,128,256},
    grid=both,
    minor grid style={draw=gray!15, line width=0.2pt},
    major grid style={draw=gray!30, line width=0.3pt},
    legend cell align=left,
    legend to name=scaling-legend,
    legend columns=3,
    legend style={
        font=\scriptsize,
        /tikz/every even column/.append style={column sep=0.7em},
        draw=black!40,
        fill=white, fill opacity=0.9, text opacity=1,
    },
    tick label style={font=\footnotesize},
    label style={font=\footnotesize},
]

\addplot[scalingIdeal]
    table[x=P, y=ideal, col sep=space]{data/strong_runtime.dat};
\addlegendentry{Ideal}

\addplot[scalingSgdFp32] table[x=P, y=sgd, col sep=space]{data/strong_runtime.dat};
\addlegendentry{SGD (FP32)}

\addplot[scalingSgdBf16] table[x=P, y=sgd_bf16, col sep=space]{data/strong_runtime.dat};
\addlegendentry{SGD (BF16)}

\addplot[scalingCfgA] table[x=P, y=ca_sgd, col sep=space]{data/strong_runtime.dat};
\addlegendentry{A (FP32)}

\addplot[scalingCfgC] table[x=P, y=ca_sgd_bf16, col sep=space]{data/strong_runtime.dat};
\addlegendentry{C (recipe)}

\addplot[scalingCfgG] table[x=P, y=ca_sgd_g, col sep=space]{data/strong_runtime.dat};
\addlegendentry{G (FP32 + TF32 TC)}

\addplot[scalingCfgD] table[x=P, y=ca_sgd_d, col sep=space]{data/strong_runtime.dat};
\addlegendentry{D (aggressive BF16)}

\addplot[scalingCfgF] table[x=P, y=ca_sgd_f, col sep=space]{data/strong_runtime.dat};
\addlegendentry{F (FP16 + FP16-accum HMMA)}

\end{loglogaxis}
\end{tikzpicture}
\caption{Strong scaling.}
\Description{Strong-scaling runtime per equivalent SGD iteration versus processor count P from 4 to 256 on log-log axes, with one marked curve per binary and an unmarked black dashed linear-scaling reference.}
\label{fig:strong-scaling}
\end{subfigure}
\hfill
\begin{subfigure}[t]{0.49\linewidth}
\centering
\begin{tikzpicture}
\begin{semilogxaxis}[
    log basis x=2,
    width=0.92\linewidth,
    height=0.62\linewidth,
    xlabel={GPUs ($P$)},
    ylabel={Weak-scaling efficiency},
    xmin=3, xmax=320,
    ymin=0, ymax=1.15,
    xtick={4,8,16,32,64,128,256},
    xticklabels={4,8,16,32,64,128,256},
    ytick={0,0.2,0.4,0.6,0.8,1.0},
    yticklabel={\pgfmathprintnumber[fixed,precision=1]{\tick}},
    grid=both,
    minor grid style={draw=gray!15, line width=0.2pt},
    major grid style={draw=gray!30, line width=0.3pt},
    tick label style={font=\footnotesize},
    label style={font=\footnotesize},
]

\addplot[scalingIdeal]
    coordinates {(4,1.0) (256,1.0)};

\addplot[scalingSgdFp32] table[x=P, y=sgd, col sep=space]{data/weak_efficiency.dat};

\addplot[scalingSgdBf16] table[x=P, y=sgd_bf16, col sep=space]{data/weak_efficiency.dat};

\addplot[scalingCfgA] table[x=P, y=ca_sgd, col sep=space]{data/weak_efficiency.dat};

\addplot[scalingCfgC] table[x=P, y=ca_sgd_bf16, col sep=space]{data/weak_efficiency.dat};

\addplot[scalingCfgG] table[x=P, y=ca_sgd_g, col sep=space]{data/weak_efficiency.dat};

\addplot[scalingCfgD] table[x=P, y=ca_sgd_d, col sep=space]{data/weak_efficiency.dat};

\addplot[scalingCfgF] table[x=P, y=ca_sgd_f, col sep=space]{data/weak_efficiency.dat};

\end{semilogxaxis}
\end{tikzpicture}
\caption{Weak scaling.}
\Description{Weak-scaling efficiency E(P) versus processor count P from 4 to 256 on semilog-x axes, with one marked curve per binary and an unmarked black dashed E=1 ideal-efficiency reference.}
\label{fig:weak-scaling}
\end{subfigure}

\vspace{0.2em}
\centerline{\ref{scaling-legend}}

\caption{Scaling of CA-SGD and SGD on NERSC Perlmutter A100 under various precision settings. Each panel includes an ideal-scaling reference (linear in (\subref{fig:strong-scaling}), $E=1$ in (\subref{fig:weak-scaling})).}
\label{fig:scaling}
\end{figure}
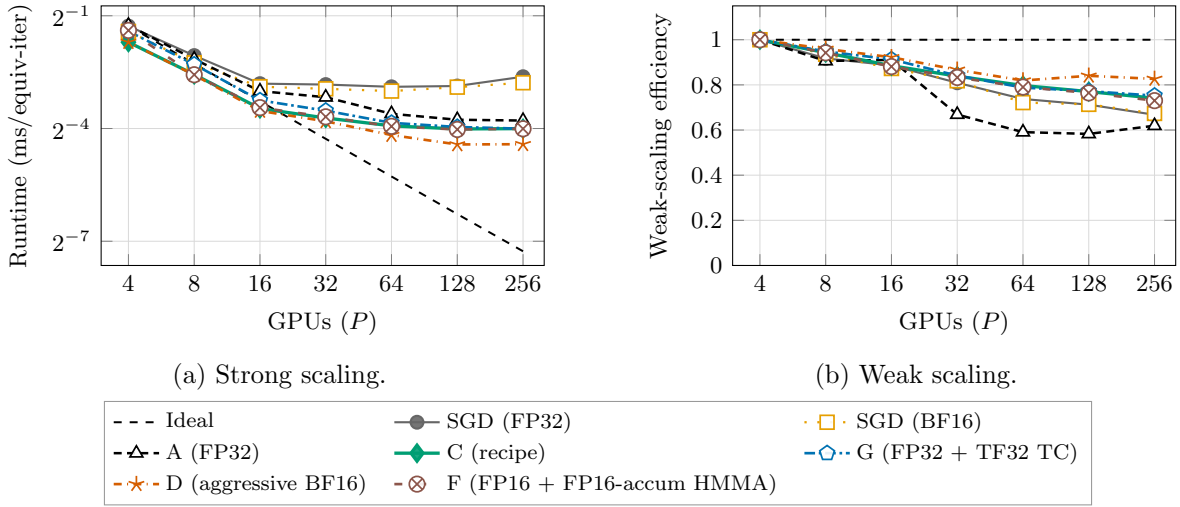

\subsection{Roofline}
\label{subsec:roofline-detail}

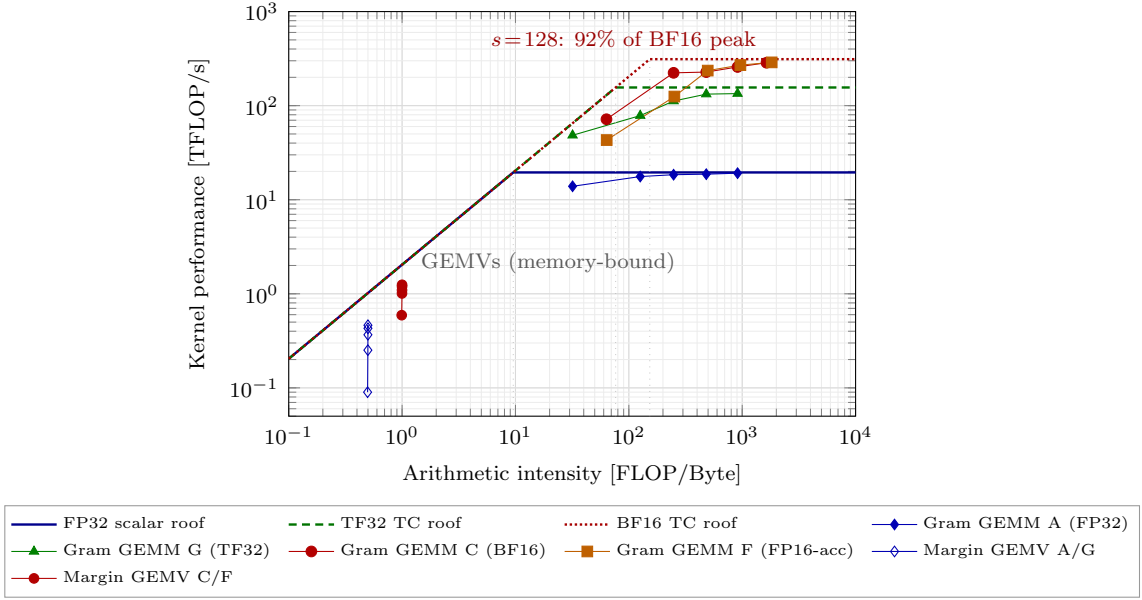
\begin{figure}[ht]
\centering
\begin{tikzpicture}
\begin{loglogaxis}[
    width=0.55\linewidth,
    height=0.42\linewidth,
    xlabel={Arithmetic intensity [FLOP/Byte]},
    ylabel={Kernel performance [TFLOP/s]},
    xmin=0.1, xmax=10000,
    ymin=0.05, ymax=1000,
    grid=both,
    minor grid style={draw=gray!15, line width=0.2pt},
    major grid style={draw=gray!30, line width=0.3pt},
    legend cell align=left,
    legend style={
        at={(0.5,-0.22)},
        anchor=north,
        legend columns=4,
        font=\tiny,
        draw=black!40,
        inner sep=2pt,
        /tikz/every even column/.append style={column sep=0.4em},
    },
    tick label style={font=\scriptsize},
    label style={font=\scriptsize},
    axis on top=false,
]

\addplot[line width=0.9pt, draw=blue!55!black]
    coordinates {(0.1, 0.2039) (9.5636, 19.5) (10000, 19.5)};
\addlegendentry{FP32 scalar roof}

\addplot[line width=0.9pt, draw=green!45!black, densely dashed]
    coordinates {(0.1, 0.2039) (76.5081, 156) (10000, 156)};
\addlegendentry{TF32 TC roof}

\addplot[line width=0.9pt, draw=red!65!black, densely dotted]
    coordinates {(0.1, 0.2039) (153.0162, 312) (10000, 312)};
\addlegendentry{BF16 TC roof}

\addplot[
    mark=diamond*, mark size=2.0pt,
    draw=blue!70!black, line width=0.4pt,
    mark options={solid, fill=blue!70!black},
] coordinates {
    (  31.9,  13.89)
    ( 126.0,  17.63)
    ( 248.2,  18.45)
    ( 481.9,  18.69)
    ( 910.2,  19.15)
};
\addlegendentry{Gram GEMM A (FP32)}

\addplot[
    mark=triangle*, mark size=2.0pt,
    draw=green!50!black, line width=0.4pt,
    mark options={solid, fill=green!50!black},
] coordinates {
    (  31.9,  48.42)
    ( 126.0,  78.39)
    ( 248.2, 111.90)
    ( 481.9, 132.58)
    ( 910.2, 134.34)
};
\addlegendentry{Gram GEMM G (TF32)}

\addplot[
    mark=*, mark size=2.0pt,
    draw=red!70!black, line width=0.4pt,
    mark options={solid, fill=red!70!black},
] coordinates {
    (  63.5,  71.70)
    ( 248.2, 223.33)
    ( 481.9, 227.42)
    ( 910.2, 257.86)
    (1638.4, 285.92)
};
\addlegendentry{Gram GEMM C (BF16)}

\addplot[
    mark=square*, mark size=2.0pt,
    draw=orange!75!black, line width=0.4pt,
    mark options={solid, fill=orange!75!black},
] coordinates {
    (  63.8,  43.08)
    ( 252.1, 124.97)
    ( 496.5, 235.86)
    ( 963.8, 269.85)
    (1820.4, 288.45)
};
\addlegendentry{Gram GEMM F (FP16-acc)}

\addplot[
    mark=diamond, mark size=2.0pt,
    draw=blue!70!black, line width=0.4pt,
] coordinates {
    (0.4961, 0.0900)
    (0.4990, 0.2517)
    (0.4995, 0.3669)
    (0.4997, 0.4630)
    (0.4998, 0.4294)
};
\addlegendentry{Margin GEMV A/G}

\addplot[
    mark=*, mark size=1.8pt,
    draw=red!70!black, line width=0.4pt,
    mark options={solid, fill=red!70!black},
] coordinates {
    (0.9921, 0.5923)
    (0.9979, 1.0086)
    (0.9989, 1.0931)
    (0.9994, 1.1968)
    (0.9996, 1.2469)
};
\addlegendentry{Margin GEMV C/F}

\node[font=\scriptsize, anchor=south east, text=red!60!black]
    at (axis cs:1638, 305)
    {$s\!=\!128$: 92\% of BF16 peak};

\node[font=\scriptsize, anchor=south west, text=black!65]
    at (axis cs:1.2, 1.3) {GEMVs (memory-bound)};

\draw[dotted, gray!55, line width=0.3pt]
    (axis cs:9.5636, 0.05)  -- (axis cs:9.5636, 19.5);
\draw[dotted, gray!55, line width=0.3pt]
    (axis cs:76.5081, 0.05) -- (axis cs:76.5081, 156);
\draw[dotted, gray!55, line width=0.3pt]
    (axis cs:153.0162, 0.05) -- (axis cs:153.0162, 312);

\end{loglogaxis}
\end{tikzpicture}
\caption{Per-kernel measured roofline at $P=1$, $m=8192$,
$n_{\rm loc}=16{,}384$, $b=32$, $H=20$. Theoretical roofs (FP32
scalar $19.5$, TF32 $156$, BF16 $312$~TFLOP/s) against $2039$~GB/s
HBM2e (knees at $9.56$, $76.5$, $153$~FLOP/B).}
\Description{Per-kernel roofline plot with arithmetic intensity on the horizontal axis and achieved throughput on the vertical axis, both log-scale, showing three compute roofs and measured points for CA-SGD Gram GEMM and margin GEMV kernels across an s sweep.}
\label{fig:roofline-combined}
\end{figure}

\Cref{fig:roofline-combined} shows a per-kernel roofline analysis on a single A100 GPU on a synthetic dataset.
We report analytical FLOP counts divided by measured kernel time for the Gram GEMM and margin GEMV kernels, which exercise different hardware instructions across the recipes in \Cref{tab:recipe-configs}.

We vary $s$ to measure kernel performance as the $s$-step length increases.
For Gram GEMM, FP32 SIMT delivers $17.6$~TFLOP/s ($90\%$ of the FP32 roof), TF32 tensor cores deliver $78.4$~TFLOP/s ($50\%$ of the TF32 roof), and BF16 with an FP32 accumulator delivers $223.3$~TFLOP/s ($72\%$ of the BF16 roof), rising to $285.9$~TFLOP/s ($92\%$ of peak) at $s=128$.
The $4\times$ jump from TF32 to BF16 at modest $s$ is the empirical reason Recipe~C uses BF16-input tensor-core GEMM.
Recipe~F uses FP16 inputs with an FP16 accumulator. At $s=16$ it sits below Recipe~C, and at $s=128$ the two converge.
The margin and gradient GEMVs sit at arithmetic intensities of $0.5$--$1.0$~FLOP/B and below $1.3$~TFLOP/s, bandwidth-bound at every precision, which matches the analytic placement of \Cref{sec:perf-model}. Their only mixed-precision benefit is a $2\times$ drop in the $Y$ reads when storage moves from FP32 to BF16.

\section{Discussion}
\label{sec:discussion}

Mixed-precision CA-SGD is effective only when the $s$-step reformulation and precision placement are designed together.
Our nine-slot error analysis justifies Recipe~C (\Cref{tab:recipe}) as the theory-covered default and gives a bounded-regime convergence neighborhood for $s$-step SGD, extending the finite-precision program for $s$-step Krylov methods~\cite{carson2014residualreplacement,carson2015sstep,carson2018adaptive,carson2022sstepmixed}.
On Perlmutter A100 GPUs, Recipe~C matches FP32 SGD while reaching up to ${\sim}16\times$ on \texttt{synth} and $5.2$--$6.2\times$ on the main datasets. Recipe~D reaches $6.8\times$ by casting the Gram AllReduce to BF16 but lies outside the verified budget.

\section*{Acknowledgments}
This work was supported by the U.S.\ Department of Energy, Office of Science, Advanced Scientific Computing Research (ASCR) under Award No.\ DE-SC0025394.
This research used resources of the National Energy Research Scientific Computing Center, a DOE Office of Science User Facility supported by the Office of Science of the U.S.\ Department of Energy under Contract No.\ DE-AC02-05CH11231, using NERSC award ASCR-ERCAP0030076.
ISM was supported by a Cornell fellowship.

\paragraph*{Generative AI Statement.}
We used Claude (Anthropic) to produce initial drafts of proofs and to develop software and experimental tooling.
The authors directed the work, accepted, rejected, or corrected all generated material, and independently verified all software, empirical results, and proofs.

\bibliographystyle{plainnat}
\bibliography{refs}

\end{document}